\documentclass{llncs}

\usepackage{amsmath,amssymb}
\usepackage{graphicx}
\usepackage{wrapfig}
\usepackage{verbatim}
\usepackage{paralist}
\usepackage{xspace}
\usepackage{algorithm}
\usepackage{algorithmic}
\usepackage{marvosym}
\usepackage{subfigure}
\usepackage[per-mode=symbol,detect-all]{siunitx}
\usepackage{graphics}
\usepackage{enumitem}
\usepackage{cite}
\usepackage{booktabs}
\usepackage[table]{xcolor}
\usepackage{pdfpages}
\usepackage{colortbl}
\usepackage{tikz,ifthen,pgfplots}
\usepackage{multirow}
\usepackage{tabularx,stackengine,collcell}

\usetikzlibrary{
  babel,
  intersections,
  arrows.meta,
  graphs,
  calc,
  positioning,
  shapes.misc,
  backgrounds,
  fit
}
\pgfplotsset{compat=1.18}

\usepackage{hyperref}
\usepackage[capitalize]{cleveref}

\newcommand{\mysubsection}[1]{\medskip\noindent\textbf{#1}}
\newcommand{\xsp}{\xspace{}}

\newcommand{\fmax}[1]{\operatorname{max}(#1)}

\newcommand{\fmin}[1]{\operatorname{min}(#1)}

\newcommand{\relu}{\mathrm{ReLU}\xsp}
\newcommand{\frelu}[1]{\relu(#1)}
\newcommand{\abs}{\mathrm{Abs}\xsp}
\newcommand{\fabs}[1]{\abs(#1)}
\newcommand{\leakyrelu}{\mathrm{LeakyReLU}\xsp}

\newcommand{\sign}{\mathrm{Sign}\xsp}
\newcommand{\maxxx}{\mathrm{Max}\xsp}
\newcommand{\bilinear}{\mathrm{Bilinear}\xsp}
\newcommand{\softmax}{\mathrm{Softmax}\xsp}
\newcommand{\sigmoid}{\mathrm{Sigmoid}\xsp}
\newcommand{\sig}[1]{{\sigma}_{#1}}

\newcommand{\veczero}{\boldsymbol{0}}
\newcommand{\vecA}{\boldsymbol{A}}
\newcommand{\vecalpha}{\boldsymbol{\alpha}}
\newcommand{\vecalphainiti}{\boldsymbol{\alpha}_{\textsc{initial}}}
\newcommand{\vecalphaselec}{\boldsymbol{\alpha}_{\textsc{select}}}
\newcommand{\vecalphagener}{\boldsymbol{\alpha}_{\textsc{gener}}}
\newcommand{\vecalphafinal}{\boldsymbol{\alpha}_{\textsc{final}}}
\newcommand{\vecbb}{\boldsymbol{b}}
\newcommand{\vecb}[1]{{\boldsymbol{b}}^{(#1)}}
\newcommand{\vecbl}[1]{{\boldsymbol{b_{\ell}}}^{(#1)}}
\newcommand{\vecbu}[1]{{\boldsymbol{b_u}}^{(#1)}}
\newcommand{\vecc}[1]{{\boldsymbol{c}}^{(#1)}}
\newcommand{\vechc}[1]{\boldsymbol{\hat{c}}^{(#1)}}
\newcommand{\vecctop}[1]{{\boldsymbol{c}}^{(#1)\top}}
\newcommand{\vechctop}[1]{\boldsymbol{\hat{c}}^{(#1)\top}}
\newcommand{\vecC}[1]{{\boldsymbol{C}}^{(#1)}}
\newcommand{\vechC}[1]{{\boldsymbol{\hat{C}}}^{(#1)}}
\newcommand{\vecdd}{\boldsymbol{d}}
\newcommand{\vecd}[1]{{\boldsymbol{d}}^{(#1)}}
\newcommand{\vecdoptim}[1]{{\boldsymbol{\mathtt{dOptimized}}}^{(#1)}}
\newcommand{\vecdbranc}[1]{{\boldsymbol{\mathtt{dBranch}}}^{(#1)}}
\newcommand{\vecdfeasi}[1]{{\boldsymbol{\mathtt{dFeasible}}}^{(#1)}}
\newcommand{\vecepsilonn}{\boldsymbol{\epsilon}}
\newcommand{\vecepsilon}[1]{{\boldsymbol{\epsilon}}^{(#1)}}
\newcommand{\vecgamma}{\boldsymbol{\gamma}}
\newcommand{\vecgammatop}{{\boldsymbol{\gamma}}^{\top}}
\newcommand{\vecl}[1]{{\boldsymbol{{\ell}}}^{(#1)}}
\newcommand{\vechl}[1]{{\boldsymbol{\hat{{\ell}}}}^{(#1)}}
\newcommand{\vecnuu}{\boldsymbol{\nu}}
\newcommand{\vechnuu}{\boldsymbol{\hat{\nu}}}
\newcommand{\vecnu}[1]{{\boldsymbol{\nu}}^{(#1)}}
\newcommand{\vechnu}[1]{\boldsymbol{\hat{\nu}}^{(#1)}}
\newcommand{\vecnutop}[1]{{\boldsymbol{\nu}}^{(#1)\top}}
\newcommand{\vechnutop}[1]{\boldsymbol{\hat{\nu}}^{(#1)\top}}
\newcommand{\vecpii}{\boldsymbol{\pi}}
\newcommand{\vecpi}[1]{{\boldsymbol{\pi}}^{(#1)}}
\newcommand{\vecpitop}[1]{{\boldsymbol{\pi}}^{(#1)\top}}
\newcommand{\vectauu}{\boldsymbol{\tau}}
\newcommand{\vectau}[1]{{\boldsymbol{\tau}}^{(#1)}}
\newcommand{\vectautop}[1]{{\boldsymbol{\tau}}^{(#1)\top}}
\newcommand{\vecu}[1]{{\boldsymbol{u}}^{(#1)}}
\newcommand{\vechu}[1]{{\boldsymbol{\hat{u}}}^{(#1)}}
\newcommand{\vecW}[1]{{\boldsymbol{W}}^{(#1)}}
\newcommand{\vecWl}[1]{{\boldsymbol{W_{\ell}}}^{(#1)}}
\newcommand{\vecWu}[1]{{\boldsymbol{W_u}}^{(#1)}}
\newcommand{\vecx}[1]{{\boldsymbol{x}}^{(#1)}}
\newcommand{\vechx}[1]{{\boldsymbol{\hat{x}}}^{(#1)}}
\newcommand{\vecxx}{\boldsymbol{x}}
\newcommand{\vechxx}{\boldsymbol{\hat{x}}}
\newcommand{\vecxtag}{\boldsymbol{x'}}
\newcommand{\score}[2]{{\mathtt{score}}^{(#1)}_{#2}}
\newcommand{\x}[2]{{x}^{(#1)}_{#2}}
\newcommand{\hx}[2]{{\hat{x}}^{(#1)}_{#2}}
\newcommand{\lowerb}[2]{{{\ell}}^{(#1)}_{#2}}
\newcommand{\lowerhb}[2]{{\hat{{\ell}}}^{(#1)}_{#2}}
\newcommand{\upperb}[2]{{u}^{(#1)}_{#2}}
\newcommand{\upperhb}[2]{{\hat{u}}^{(#1)}_{#2}}
\newcommand{\pmnr}{\textsc{Pmnr}}
\newcommand{\pmnrall}{\textsc{PmnrAll}}
\newcommand{\stopCondition}{\textsc{StopCondition}}
\newcommand{\pickAlphas}{\textsc{PickAlphas}}
\newcommand{\singleNeuronTightening}{\textsc{SingleNeuronTightening}}
\newcommand{\selectNeurons}{\textsc{SelectNeurons}}
\newcommand{\generatePMNR}{\textsc{GeneratePMNR}}
\newcommand{\PostTighten}{\textsc{PostTighten}}
\newcommand{\Branches}{\textsc{Branches}}

\newcommand{\initialPMNR}{\textsc{initialPMNR}}
\newcommand{\branchCombinations}{\textsc{BranchCombinations}}
\newcommand{\optimizePMNR}{\textsc{OptimizePMNR}}
\newcommand{\concreteBounds}{\mathtt{ConcreteB}}
\newcommand{\singleNeuronBounds}{\mathtt{SingleB}}
\newcommand{\neurons}{\mathtt{Neurons}}
\newcommand{\multiNeuronBounds}{\mathtt{MultiB}}
\newcommand{\branchPoints}{\mathtt{BranchPoints}}
\newcommand{\branchBounds}{\mathtt{BranchB}}
\newcommand{\infeasibleBranches}{\mathtt{InfeasibleBranches}}
\newcommand{\branchCombination}{\mathtt{branchCombination}}

\newcommand{\ReluSignMax}{\textsc{ReluSignMax}\xsp}
\newcommand{\ReluBilinearSoftmax}{\textsc{ReluBilinearSoftmax}\xsp}
\newcommand{\LeakyReluFive}{\textsc{LeakyRelu}\(5\times100\)\xsp}
\newcommand{\LeakyReluEight}{\textsc{LeakyRelu}\(8\times100\)\xsp}
\newcommand{\LeakyReluFourteen}{\textsc{LeakyRelu}\(14\times28\)\xsp}
\newcommand{\LeakyReluSigmoid}{\textsc{LeakyReluSigmoid}\xsp}
\newcommand{\MNIST}[1]{\mathtt{MNIST}_{#1}\xsp}
\newcommand{\FBC}{\texttt{F+BC} }
\newcommand{\Ra}{\Rightarrow}
\newcommand{\supth}[1]{{#1}\textsuperscript{th}}

\newcommand{\R}[1]{{\mathbb{R}}^{#1}}
\newcommand{\SAT}{\mathtt{SAT}}
\newcommand{\UNSAT}{\mathtt{UNSAT}}
\newcommand{\Din}{{\mathcal{D}}_{\mathtt{in}}}
\newcommand{\Dout}{{\mathcal{D}}_{\mathtt{out}}}
\newcommand{\mpos}[1]{{[#1]}_{+}}
\newcommand{\mneg}[1]{{[#1]}_{-}}
\newcommand{\bigmpos}[1]{{\left[#1\right]}_{+}}
\newcommand{\bigmneg}[1]{{\left[#1\right]}_{-}}

\begin{document}

\title{
  Neural Network Verification using \\
  Partial Multi-Neuron Relaxation
}
\titlerunning{Partial Multi-Neuron Relaxation}
\author{
  Ido Shmuel \orcidID{0009-0002-9870-549X} \and
  Guy Katz\orcidID{0000-0001-5292-801X}
}
\authorrunning{I. Shmuel, G. Katz}
\institute{
  The Hebrew University of Jerusalem, Jerusalem, Israel \\
  \email{ \{ido.shmuel, g.katz\}@mail.huji.ac.il}
}

\maketitle
\begin{abstract}
	The increasing integration of deep neural networks in critical
	systems has spawned a theoretical and practical interest in
	formally guaranteeing safety properties about their behavior.
	To achieve this, contemporary verification algorithms rely on
	computing linear relaxations for a network's non-linear
	activation functions. Existing approaches for linear relaxations
	typically fall into one of two categories: single-neuron
	relaxation, in which each activation neuron is bounded in terms
	of its sources; and multi-neuron relaxation, in which linear
	bounds involving multiple activation neurons and their sources
	are calculated. However, existing methods might fail to balance
	tightness and scalability, as single-neuron bounds might not
	derive sufficiently tight bounds necessary for verification to
	complete, whereas generating multi-neuron relaxation for all
	activation neurons is computationally expensive. In this paper,
	we present a middle-ground approach featuring \emph{partial
	multi-neuron relaxation}, in which we generate multi-neuron
	bounds for only a small, heuristically selected subset of
	neurons. To achieve this, we build upon existing branching
	heuristics for selecting neurons and for optimizing bounding
	hyper-planes for multi-neuron bounds. We integrated our proposed
	method within the Marabou verifier, and obtained favorable
	results in comparison to existing bound tightening methods. Our
	experiments showcase the potential of our technique for neural
	network verification.
\end{abstract}

\section{Introduction}\label{sec:introduction}

Deep neural networks~\cite{GoodBengCour16} (DNNs) are increasingly
being adopted as key components in mission-critical systems. They
have been achieving unprecedented performance in diverse domains such
as natural language processing~\cite{Gemini2025}, protein structure
prediction~\cite{AlphaFold2021,ESMFold2023}, image
recognition~\cite{SiZi14}, medical analysis~\cite{CaWaChJiZhTiWa23},
aircraft collision avoidance~\cite{JuLoBrOwKo16},
self-driving~\cite{BoDeDwFiFlGoJaMoMuZhZhZhZi16}, and task
scheduling~\cite{PeBaChWuMeLi19},
often greatly improving upon the results of algorithms crafted by
human experts.

However, despite their immense success, the opacity of DNNs raises
new concerns regarding their stability and reliability. Unlike
classical, human-crafted algorithms, DNNs are perceived as
black-boxes, making it troublesome to correctly reason about their
decision making~\cite{BuHu21}. Moreover, DNNs are known to be
susceptible to adversarial
perturbations~\cite{SzZaSuBrErGoFe13,GoShSz15,ChWuGuLiJh17}, where
low-magnitude changes to their inputs result in incorrect outputs.
The absence of a formal proof of correctness of DNNs might raise
doubts about the robustness of existing applications of DNNs and
potentially slow down their adoption.

In order to address these issues, diverse strategies have been
suggested to feasibly solve the problem of formal verification of
DNNs~\cite{LiXiLi23}. Despite recent
advances~\cite{KaLaBaBrDuFlJoKoMaNgWu25}, these methods
have limited scalability, in light of the fact DNN verification has
been proven to be NP-complete for DNNs with piecewise-linear
activation functions~\cite{KaBaDiJuKo17}.

A major component in existing methods for DNN verification is the
branch-and-bound (BaB) paradigm~\cite{BuLuTuToKoKu20}. Due to the
non-linear nature of activation functions, DNN verification often
necessitates performing recursive case splitting, resulting in a
large search space whose size grows exponentially as a function of
the number of activation neurons. Under the BaB paradigm, prior to
splitting the original verification problem to sub-problems,
verifiers take advantage of bound tightening tactics to
deduce upper and lower bounds on the values of the networks' neurons.
This often results in a major size reduction of the search space,
facilitating verification.

In order to derive bounds on a DNN's neurons, solvers leverage linear
relaxations of a network's non-linear activation functions. By
calculating linear over-approximations on activation neurons as a
function of their sources, solvers may leverage technique
featuring symbolic bound propagation~\cite{WaPeWhYaJa18,GaGePuVe19}
and Linear Programming (LP)~\cite{WaPiWhYaJa18,SaYaZhHsZh19} to
quickly gather bounds on the network's neurons. Existing
methods typically fall into one of two categories: single-neuron
relaxation~\cite{GaGePuVe19,ZhWeChHsDa18}, in which each of the
calculated linear over-approximations involve a single activation
neuron; and multi-neuron relaxation, which features linear bounds
consisting of multiple activation
neurons~\cite{SiGaPuVe19,FeNuJoVe22,MuMaSiPuVe22,SuChZiKrHu23}.
Although the former strategy is highly scalable, it
might result in insufficiently tight over-approximations, due to the
convex relaxation barrier~\cite{SaYaZhHsZh19}. In contrast, the
latter approach results in tighter bounds while incurring a
higher computational toll.

In this paper, we present a novel approach to bound tightening, which
seeks a better balance between scalability and tightness. We propose
a general framework for performing \emph{partial multi-neuron
relaxation} (PMNR), namely, generating multi-neuron over-
approximations for only a heuristically selected subset of neurons,
while using tunable single-neuron bounds for the remainder of the
network. Our approach can be instantiated with various heuristics for
selecting neurons and hyper-planes for multi-neuron relaxations, and
we demonstrate that multiple existing branching heuristics can be
used for this purpose. Whereas in the BaB paradigm, branching
heuristics indicate a single neuron whose bounds may be improved by
first performing a case-split, in our approach we use these
heuristics to identify cases where it is beneficial to infer a
multi-neuron bound.

We implemented PMNR within the popular Marabou verification
tool~\cite{KaHuIbJuLaLiShThWuZeDiKoBa19,MarabouII2025}. We evaluated
our implementation on local robustness queries with fully connected
DNNs, trained on the MNIST dataset~\cite{LeBeHa98}; and compared it
to other tightening algorithms implemented in Marabou. We discovered
that the PMNR-enhanced Marabou solved \(49\%\) more queries than base
Marabou, with a runtime reduction of \(17\%\) on our benchmarks.
Our experiments demonstrate the significant potential of the PMNR
approach for bound tightening.
Our implementation is available online~\cite{ShKa2026Code}.

The rest of the paper is organized as follows. In
Section~\ref{sec:background} we provide the necessary background on
DNNs, their verification, and linear relaxations of activation
functions. Next, in Section~\ref{sec:pmnr-paradigm} we describe our
general technique for verifying DNNs using partial multi-neuron
relaxations, before instantiating it with specific heuristics in
Section~\ref{sec:instantiating-pmnr}. Next, we evaluate the
performance of our approach in
Section~\ref{sec:experiments-and-evaluation}. We cover related work
in Section~\ref{sec:related-work}, and conclude with
ideas for future research in
Section~\ref{sec:conclusion-and-future-work}.

\section{Background}\label{sec:background}

\subsection{Deep Neural Networks and their Verification}\label{background:subsec:dnns-and-verification}
\mysubsection{Deep Neural Networks.} A deep neural network (DNN)
\(N:\R{n_0} \to \R{n_L}\) contains an input layer, \(L-1\) hidden
layers, and an output layer. The \supth{\(i\)} layer of the network
is a real-valued vector of size \(n_i\). We denote the \supth{\(i\)}
layer as \(\vechx{i}\) and its \supth{\(j\)} neuron as \(\hx{i}{j}\).
For \(0 < i < L\), the DNN's hidden layers are iteratively computed
given the recursion formula
\(\vechx{i} = \sig{i} ( \vecW{i} \vechx{i-1} + \vecb{i} )\), and
the output layer is defined as
\(\vecx{L} = \vecW{L} \vechx{L-1} + \vecb{L}\). For each \(i\),
\(\vecW{i}\) is a weight matrix, \(\vecb{i}\) is a bias vector, and
\(\sig{i} : \R{n_i} \to \R{n_i}\) is a (usually non-linear)
activation function. Note that \(N\)'s output
\(N(\vechx{0}) = \vecx{L}\) does not contain post-activation values,
which could be modeled as setting \(\sig{L}\) to be the identity
function. We also define pre-activation values with the relation
\(\vecx{i} = \vecW{i} \vechx{i-1} + \vecb{i}\).
Since the activations \(\sig{i}\) are modeled as arbitrary
multivariate functions, our formulation generalizes to several common
DNN architectures (e.g. fully-connected, convolutional, residual)
equipped with arbitrary activation functions.

\mysubsection{Example Network.} Consider the DNN in
Fig.~\ref{fig:example-network} with input
\(\vechx{0} \in [-1, 1]^2\), one hidden layer containing
\(\fabs{x} = |x|\) (absolute value) activations with pre-activation
\(\vecx{1}\) and post-activation \(\vechx{1}\), another hidden layer
containing two \(\frelu{x} = \fmax{0, x}\) activations with
pre-activation \(\vecx{2}\) and post-activation \(\vechx{2}\), and a
single output \(\x{3}{0}\). Neurons
\(\hx{1}{0}, \hx{2}{1}, \hx{3}{0}\) have bias values of \(1, 2, 26.1\)
respectively, and all other neurons have a bias of \(0\). The
piecewise-linear functions \(\abs\) and \(\relu\) are applied
element-wise for multi-dimensional inputs. The network's neurons are
computed recursively:

\begin{figure}[ht]
	\centering
	\begin{tikzpicture}[>=Stealth,
			node distance=10ex,
			text height=1.5ex,text depth=.25ex,
			invis/.style={circle, minimum size=0ex},
			input/.style={circle, minimum size=5ex,
					thick, draw=blue, fill=white},
			hidden/.style={circle, minimum size=5ex,
					thick, draw=black!50!white, fill=white},
			output/.style={circle, minimum size=5ex,
					thick, draw=red, fill=white},
			background/.style={rounded corners,
			            fill=blue!10!white, draw=black}]

		\node (x0_0)   [input]                                     {$\hx{0}{0}$};
		\node (x0_1)   [input, below=15ex of x0_0]                 {$\hx{0}{1}$};
		\node (x1_0)   [hidden, right=15ex of x0_0, yshift=10ex]   {$\hx{1}{0}$};
		\node (x1_1)   [hidden, right=15ex of x0_0, yshift=-10ex]  {$\hx{1}{1}$};
		\node (x1_2)   [hidden, right=15ex of x0_1, yshift=-10ex]  {$\hx{1}{2}$};
		\node (x2_0)   [hidden, right=15ex of x1_0, yshift=-10ex]  {$\hx{2}{0}$};
		\node (x2_1)   [hidden, right=15ex of x1_2, yshift=10ex]   {$\hx{2}{1}$};
		\node (x3_0)   [output, right=15ex of x2_0, yshift=-10ex]  {$\x{3}{0}$};

		\draw [->] (x0_0) -- (x1_0)    node[midway, above]                         {1};
		\draw [->] (x0_0) -- (x1_1)    node[midway, below]                         {2};
		\draw [->] (x0_1) -- (x1_1)    node[midway, above]                         {-3};
		\draw [->] (x0_1) -- (x1_2)    node[midway, below]                         {1};

		\draw [->] (x1_0) -- (x2_0)    node[midway, above]                         {1};
		\draw [->] (x1_0) -- (x2_1)    node[near start, above right, yshift=-1ex]  {-1};
		\draw [->] (x1_1) -- (x2_0)    node[very near start, above]                {1};
		\draw [->] (x1_1) -- (x2_1)    node[very near start, below]                {1};
		\draw [->] (x1_2) -- (x2_0)    node[near start, below right, yshift=1ex]   {-1};
		\draw [->] (x1_2) -- (x2_1)    node[midway, below]                         {-5};

		\draw [->] (x2_0) -- (x3_0)   node[midway, above]                          {-1};
		\draw [->] (x2_1) -- (x3_0)   node[midway, below]                          {-3};

		\node (x1_0_bias)   [align=center, below=0.5ex of x1_0]   {1};
		\node (x1_1_bias)   [align=center, below=0.5ex of x1_1]   {0};
		\node (x1_2_bias)   [align=center, below=0.5ex of x1_2]   {0};

		\node (x2_0_bias)   [align=center, below=0.5ex of x2_0]   {0};
		\node (x2_1_bias)   [align=center, below=0.5ex of x2_1]   {2};

		\node (x3_0_bias)   [align=center, below=0.5ex of x3_0]   {26.1};

		\node (x0_0_range)   [align=center, above=0.5ex of x0_0]  {$[-1, 1]$};
		\node (x0_1_range)   [align=center, above=0.5ex of x0_1]  {$[-1, 1]$};

		\node (x1_0_range)   [align=center, above=0.5ex of x1_0]  {$[0, 2]$};
		\node (x1_1_range)   [align=center, above=0.5ex of x1_1]  {$[0, 5]$};
		\node (x1_2_range)   [align=center, above=0.5ex of x1_2]  {$[0, 1]$};

		\node (x2_0_range)   [align=center, above=0.5ex of x2_0]  {$[0, 4]$};
		\node (x2_1_range)   [align=center, above=0.5ex of x2_1]  {$[0, 6]$};

		\node (x3_0_range)   [align=center, above=0.5ex of x3_0]  {$[12.1, 26.1]$};

	\end{tikzpicture}
	\caption{An example neural network.}
	\label{fig:example-network}
\end{figure}
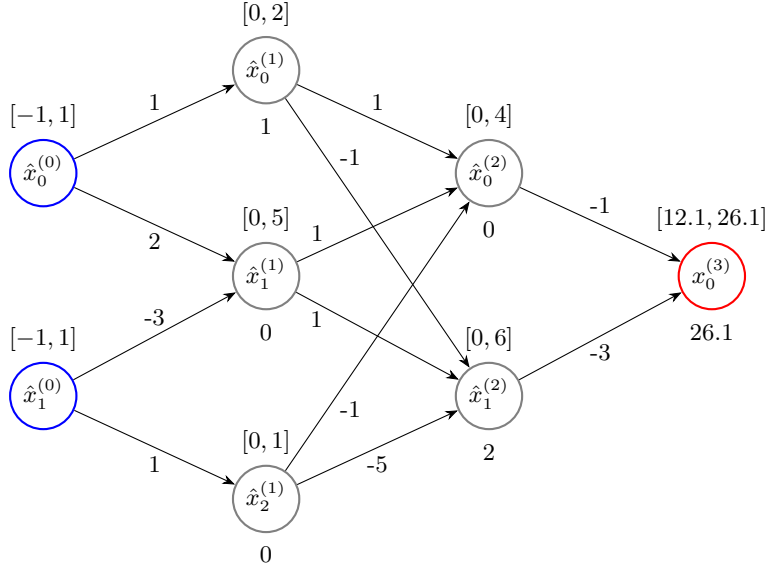
\[ \vecx{1} = \begin{bmatrix} 1 \enspace& 0 \\ 2 \enspace& -3 \\ 0 \enspace& 1\end{bmatrix} \vechx{0} + \begin{bmatrix} 1 \\ 0 \\ 0 \end{bmatrix},\quad \vechx{1} = \fabs{\vecx{1}}. \]
\[ \vecx{2} = \begin{bmatrix} 1 \enspace& 1 \enspace& -1 \\ 1 \enspace& -1 \enspace& -5\end{bmatrix} \vechx{1} + \begin{bmatrix} 0 \\ 2 \end{bmatrix},\quad \vechx{2} = \frelu{\vecx{2}}. \]
\[ \vecx{3} = \begin{bmatrix} -1 \enspace& -3 \end{bmatrix} \vechx{2} + 26. \]
Given input \(\vechx{0} = (0.4, -0.6)^{\top}\), \(N\)'s hidden
neurons have these values:
\[\vechx{1} = (\fabs{1.4}, \fabs{2.6}, \fabs{-0.6})^{\top} = (1.4, 2.6, 0.6)^{\top}.\]
\[\vechx{2} = (\frelu{3.4}, \frelu{0.2})^{\top} = (3.4, 0.2)^{\top}.\]
The output of the network is \(N(\vechx{0}) = \vecx{3} = 22.1\).

\mysubsection{Neural Network Verification.} Neural network
verification~\cite{JaBaKa20} is the process of soundly deciding
whether a safety property holds in a neural network's output given
known bounds on its inputs. Formally, a verification query is a
triple \(Q = (N, \Din, \Dout)\), consisting of a DNN \(N\), an input
domain \(\Din \subset \R{n_0}\) and an output domain
\(\Dout \subset \R{n_L}\), which typically represents an unsafe
behavior of \(N\). DNN verification is framed as a satisfiability
problem: \(Q\) is satisfiable (\(\SAT\)) if there exists \(\vecxx\)
for which the predicate
\(\vecxx \in \Din \wedge N(\vecxx) \in \Dout\) holds (i.e. \(N\)
demonstrates undesirable behavior \(\Dout\)); otherwise, it is
considered unsatisfiable (\(\UNSAT\)). Any common verification query
could be easily rewritten~\cite{ElGoKa20,IsZoBa23} into a query with
a single-output DNN whose output domain is \(\Dout = (0, \infty)\).
We will therefore assume for the remainder of this paper that all
queries are of this simplified form unless indicated otherwise.

As an example, consider the verification query composed of the DNN in
Fig.~\ref{fig:example-network}, the input domain \([-1, 1]^2\) and
the output domain \((0, \infty)\). The input
\(\vechx{0} = (0.4, -0.6)^{\top}\) satisfies it because
\(N(\vechx{0}) = 22.1 > 0\). Hence, \((0.4, -0.6)^{\top}\) is
considered a satisfying assignment for this query, and no sound
verifier would consider it unsatisfiable.

\subsection{Branch and Bound (BaB)}\label{background:subsec:branch-and-bound}
Branch-and-bound (BaB) is a key technique, applied by various neural
network verifiers~\cite{BuLuTuToKoKu20,WaZhXuLiJaHsKo21,FeNuJoVe22}.
It is formed by interleaving calls to a bound tightening algorithm
which calculates concrete bounds
\(\vechl{i} \le \vechx{i} \le \vechu{i},\quad \vecl{i} \le \vecx{i}
\le \vecu{i}\) for a DNN's neurons, and a branching method which
recursively splits the verification problem into smaller,
easier-to-solve sub-problems, giving rise to a search tree of an
exponentially increasing size. The original problem is determined to
be satisfiable if and only if at least one sub-problem is declared
\(\SAT\) by the verifier. Bound tightening is invoked post-splitting
to limit the growth rate of the search tree. In order to generate
sub-problems, verifiers commonly use case splitting on activation
neurons~\cite{WaZhXuLiJaHsKo21}. Case splitting is typically applied
for piecewise-linear activation neurons, which are then split into a
collection of linear constraints, one for each linear segment; though
it has been applied successfully for general activation functions as
well~\cite{ShJiKoJaHsZh25}. Given the massive scale of branching
trees in practice, verifiers heavily leverage heuristics and other
techniques to prune infeasible sub-problems and limit the number of
sub-problems that are generated as a result of case splitting.

\subsection{Single-Neuron Relaxation}\label{background:subsec:single-neuron-relaxation}
\mysubsection{Linear Over-Approximations.} Contemporary bound
tightening algorithms reason about a DNN's general non-linearities by
replacing them with linear
over-approximations~\cite{ZhWeChHsDa18,GaGePuVe19,LiXiLi23}. Namely,
a DNN's sound single-neuron relaxation is a collection of linear
bounds of the form:
\[ \vecWl{i} \vecx{i} + \vecbl{i} \le \sig{i}(\vecx{i}) = \vechx{i} \le \vecWu{i} \vecx{i} + \vecbu{i} \]
These bounds are \emph{sound} if they hold
whenever \(\vecl{i} \le \vecx{i} \le \vecu{i}\). The symbolic weight
matrices \(\vecWl{i}, \vecWu{i}\) and symbolic bias vectors
\(\vecbl{i}, \vecbu{i}\) are chosen as a function of
\(\sig{i}, \vecl{i}, \vecu{i}\) to guarantee soundness using known
methods~\cite{GaGePuVe19,XuZhWaWaJaLiHs20,PaWa22}. They may depend
on an external, tunable parameter
\(\vecalpha\)~\cite{XuZhWaWaJaLiHs20}.

For example, consider again the \(\relu, \abs\) activation functions
from the network in Fig.~\ref{fig:example-network}, which are
piecewise-linear with two linear phases. Given known bounds
\(\x{i}{j} \in [\lowerb{i}{j}, \upperb{i}{j}]\),
sound linear bounds on \(\frelu{\x{i}{j}}, \fabs{\x{i}{j}}\) are:
\begin{align*}
	 & \begin{cases}
		   0 \leq \frelu{\x{i}{j}} \leq 0                                                                                            & \text{if\ } \upperb{i}{j}\leq 0 \\
		   \x{i}{j} \leq \frelu{\x{i}{j}} \leq \x{i}{j}                                                                              & \text{if\ } \lowerb{i}{j}\geq 0 \\
		   \alpha \x{i}{j} \leq \frelu{\x{i}{j}} \leq \frac{\upperb{i}{j} (\x{i}{j} - \lowerb{i}{j})}{\upperb{i}{j} + \lowerb{i}{j}} &
		   \text{otherwise, for any\ } \alpha \in [0, 1]                                                                                                               \\
	   \end{cases} \\
	 & \begin{cases}
		   -\x{i}{j} \leq \fabs{\x{i}{j}} \leq -\x{i}{j}                                 & \text{if\ } \upperb{i}{j}\leq 0 \\
		   \x{i}{j} \leq \fabs{\x{i}{j}} \leq \x{i}{j}                                   & \text{if\ } \lowerb{i}{j}\geq 0 \\
		   \alpha \x{i}{j} \leq \fabs{\x{i}{j}} \leq \fmax{\lowerb{i}{j}, \upperb{i}{j}} &
		   \text{otherwise, for any\ } \alpha \in [-1, 1]
	   \end{cases}
\end{align*}\label{eqn:relu-abs-linear-relaxation}

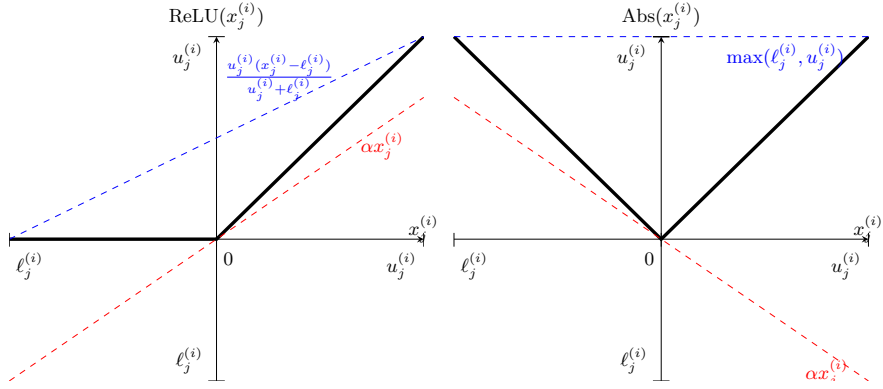
\begin{figure}[ht]
	\centering
	\begin{tikzpicture}[scale=0.8]
		\begin{axis}[domain=-1:1,
				axis lines=middle,
				ticks=none,
				x label style={at={(axis description cs:1,0.5)},
				               anchor=north},
				y label style={at={(axis description cs:0.5,1)},
				               anchor=south},
				xlabel=$\x{i}{j}$,
				ylabel=$\relu(\x{i}{j})$,]
			\addplot[color=black, ultra thick,] {(x>=0) * x + (x<0) * 0)} node[below right, pos=0.7, yshift=-2ex] {};
			\addplot[blue, dashed] {0.5*(x+1)} node[left, pos=0.8] {$\frac{\upperb{i}{j} (\x{i}{j} - \lowerb{i}{j})}{\upperb{i}{j} + \lowerb{i}{j}}$};
			\addplot[red, dashed] {0.7*x} node[below, pos=0.9] {$\alpha \x{i}{j}$};
			\draw (axis description cs:0, 0.43)   -- (axis description cs:0, 0.39) node[below right, pos=0.9]   {$\lowerb{i}{j}$};
			\draw (axis description cs:0.5, 0.43) -- (axis description cs:0.5, 0.39) node[below right, pos=0.9] {$0$};
			\draw (axis description cs:1, 0.43)   -- (axis description cs:1, 0.39) node[below left, pos=0.9]    {$\upperb{i}{j}$};
			\draw (axis description cs:0.52, 0)   -- (axis description cs:0.48, 0) node[above left, pos=0.9]    {$\lowerb{i}{j}$};
			\draw (axis description cs:0.52, 1)   -- (axis description cs:0.48, 1) node[below left, pos=0.9]    {$\upperb{i}{j}$};
		\end{axis}
		\hfill
	\end{tikzpicture}
	\begin{tikzpicture}[scale=0.8]
		\begin{axis}[domain=-1:1,
				axis lines=middle,
				ticks=none,
				x label style={at={(axis description cs:1,0.5)},
				               anchor=north},
				y label style={at={(axis description cs:0.5,1)},
				               anchor=south},
				xlabel=$\x{i}{j}$,
				ylabel=$\fabs{\x{i}{j}}$,]
			\addplot[color=black, ultra thick,] {(x>=0) * x + (x<0) * -x)} node[below right, pos=0.7, yshift=-2ex] {};
			\addplot[blue, dashed] {1} node[below, pos=0.8] {$\max(\lowerb{i}{j}, \upperb{i}{j})$};
			\addplot[red, dashed] {-0.7*x} node[below, pos=0.9] {$\alpha \x{i}{j}$};
			\draw (axis description cs:0, 0.43)        -- (axis description cs:0, 0.39) node[below right, pos=0.9]  {$\lowerb{i}{j}$};
			\draw (axis description cs:0.5, 0.43)      -- (axis description cs:0.5, 0.39) node[below left, pos=0.9] {$0$};
			\draw (axis description cs:1, 0.43)        -- (axis description cs:1, 0.39) node[below left, pos=0.9]   {$\upperb{i}{j}$};
			\draw (axis description cs:0.52, 0)        -- (axis description cs:0.48, 0) node[above left, pos=0.9]   {$\lowerb{i}{j}$};
			\draw[thick] (axis description cs:0.52, 1) -- (axis description cs:0.48, 1) node[below left, pos=0.9]   {$\upperb{i}{j}$};
		\end{axis}
	\end{tikzpicture}
	\caption{An illustration of linear bounds on the $\relu$ and
	         $\abs$ functions over
			 $x\in[\lowerb{i}{j},\upperb{i}{j}]$ where
			 $\lowerb{i}{j}<0<\upperb{i}{j}$. $\alpha \x{i}{j}$ is a
			 sound linear lower bound on $\frelu{\x{i}{j}}$ for
			 $\alpha \in [0, 1]$, and it is a sound linear lower
			 bound on $\fabs{\x{i}{j}}$ for $\alpha \in [-1, 1]$.}
	\label{fig:linear-bounds}
\end{figure}

In the first two cases of the inequalities above, the two activation
functions are known to have a fixed phase in the domain
\(\x{i}{j} \in [\lowerb{i}{j}, \upperb{i}{j}]\), and their linear
bounds are identical to their respective linear phases. Else, they
are said to have an unfixed phase in the given domain, and they are
relaxed into a pair of linear constraints. The linear relaxations in
the unfixed case are illustrated in Fig.~\ref{fig:linear-bounds}.

\mysubsection{Symbolic Bound Tightening (SBT).}
Symbolic Bound Tightening is a common, light-weight tightening method
requiring linear over-approximations on a DNN's non-linear
activations. In this method, linear bounds of every neuron in the
network as a function the previous layer's neurons are computed
iteratively using back-substitution and concretization. Two
noteworthy variations of SBT are Symbolic
Intervals~\cite{WaPeWhYaJa18} and DeepPoly~\cite{GaGePuVe19}.

\mysubsection{Linear Programming (LP).}
Alternative approaches to bound tightening reduce the DNN
verification query \(Q\) to a linear program. The existence a of
positive solution to the following LP establishes the satisfiability
of \(Q\):
\begin{align*}
	\min_{\vecalpha, \vecxx \in \Din} \qquad & N(\vecxx) = \vechx{L}                        \\
	s.t. \quad                               & \vecx{i} = \vecW{i} \vechx{i-1} + \vecb{i}   \\
	                                         & \vechx{i} \ge \vecWl{i} \vecx{i} + \vecbl{i} \\
	                                         & \vechx{i} \le \vecWu{i} \vecx{i} + \vecbu{i}
\end{align*}\label{eqn:relaxed-lp}

This LP is typically dispatched by an LP solver or duality-based
strategies~\cite{Eh17,SaYaZhHsZh19}. Alternately, the problem might
be transformed into a Mixed-Integer LP (MILP) instance, and then
solved by invoking a MILP solver~\cite{TjXiTe17}.

\subsection{Multi-Neuron Relaxation}\label{background:subsec:multi-neuron-relaxation}
Although single-neuron relaxation allows for scalable bound
tightening, its precision is inherently limited by the convex
relaxation barrier, even for simple ReLU
activations~\cite{SaYaZhHsZh19}. A multi-neuron bound is a linear
bound involving multiple activations and their associated
pre-activation values. Formally, given activation neurons
\(\neurons = \{\hx{i_1}{j_1}, \ldots \hx{i_d}{j_d}\}\), multi-neuron
linear bounds are a bounding polyhedron of the form
\(\sum_{k=1}^{d} ( \vecc{k} \hx{i_k}{j_k} - \vecC{k} \vecx{i_k} ) \le
\vecdd\). The technique of multi-neuron relaxation bypasses the
convex barrier by incorporating multi-neuron linear bounds.  Some
existing verification tools apply this technique, and are able to
learn stronger bounds by calculating multi-neuron bounds for all
neurons~\cite{SiGaPuVe19,FeNuJoVe22,MuMaSiPuVe22}.

\section{Partial Multi-Neuron Relaxation Paradigm}\label{sec:pmnr-paradigm}

While the technique of Multi-Neuron relaxation produces tighter
bounds compared to Single-Neuron Tightening, it incurs a significant
runtime overhead. Current methods~\cite{SiGaPuVe19,MuMaSiPuVe22}
either require calculating multi-neuron bounds for every activation
neuron in a given DNN, solving MILPs~\cite{ZhWaLiLiJaHsKo22}, or
having already performed branching~\cite{ZhBrHaZh24}. The first two
might not scale well for larger networks, and the third does not
apply for initial (pre-branching) bound tightening.

In order to achieve more accurate bounds compared to Single-Neuron
Tightening, while avoiding the higher cost of Multi-Neuron
Tightening, we propose to extend single-neuron relaxation by
heuristically selecting a small subset of neurons and generating
multi-neuron bounds only for these neurons. This allows to circumvent
the convex relaxation barrier without needing to calculate
multi-neuron bounds for all activation neurons.

Though it is difficult to know \textit{a priori} which subset of
neurons will yield the tightest bounds, we argue that existing
branching heuristics might be suitable for the task of neuron
selection. The motivation is that these heuristics are already
designed to identify neurons for which Single-Neuron Relaxation fails
to produce sufficiently tight bounds, so that branching can be
performed on them. Here, instead of performing branching, we propose
to tighten these neurons' bounds by incorporating them in
multi-neuron bound calculation.

In this section, we introduce the concept of lemmas, and outline our
Partial Multi-Neuron Relaxation (PMNR) bound tightening framework ---
the pseudo-code of which appears in Algorithm~\ref{alg:pmnr}. Next,
we describe each step in greater detail and prove soundness
properties.

\mysubsection{Derived Lemmas.} For the purpose of defining soundness,
we introduce the following definitions pertaining to lemmas inferred
from a verification query or from other lemmas.

\begin{definition}\label{def:query-derived-lemmas}
The set of constraints \(C\) on \(\vechx{i}, \vecx{i}\) is a lemma
derived from verification query \(Q\) if
\(\vechx{0} \in \Din \wedge N(\vechx{0}) \in \Dout\) implies
\(\vechx{i}, \vecx{i}\) satisfy all the constraints in \(C\), and we
denote \(Q \Ra C\).
\end{definition}

\begin{definition}\label{def:lemma-derived-lemmas}
For two sets \(C_1, C_2\) of constraints on \(\vechx{i}, \vecx{i}\),
\(C_2\) is a lemma derived from \(C_1\) if any
\(\vechx{i}, \vecx{i}\) which satisfy \(C_1\)'s constraints also
satisfy \(C_2\)'s, and we denote \(C_1 \Ra C_2\).
\end{definition}

As a corollary of Definition~\ref{def:query-derived-lemmas}, if
\(Q \Ra C\) and for no choice of \(\vechx{0} \in \R{n_0}\) do
\(\vechx{i}, \vecx{i}\) satisfy \(C\)'s constraints, \(Q\) is
necessarily unsatisfiable.

\mysubsection{Single-Neuron Relaxation.} Our framework uses as a
backend a single-neuron relaxation-based bound tightening method. Any
number of existing methods can be plugged in for this purpose, and we
invoke them through a call to the abstract
\(\singleNeuronTightening\) method --- which returns concrete bounds
\(\concreteBounds\), single-neuron relaxation \(\singleNeuronBounds\)
and a optimizable parameter \(\vecalphainiti\). Here,
\(\singleNeuronBounds\) is a set of linear over-approximations (as in
Subsection~\ref{background:subsec:single-neuron-relaxation}) which
depend on an optimizable parameter \(\vecalpha\), and
\(\singleNeuronBounds(\vecalpha)\) is the resulting linear
over-approximations by substituting a value of \(\vecalpha\). Our
framework requires \(\concreteBounds\) and
\(\singleNeuronBounds(\vecalpha)\) returned by
\(\singleNeuronTightening\) are lemmas learned from \(Q\) for all
values of \(\vecalpha\).

\mysubsection{Partial Multi-Neuron Relaxation.} In the case
\(\concreteBounds\) does not contain any constraint which is a
contradiction (which we denote by \(\bot\)), we proceed to
calculating multi-neuron bounds for a heuristically selected subset
of neurons. The process is repeated until \(\stopCondition\) becomes
true or \(\concreteBounds\) contains a contradiction \(\bot\).

First, at line~\ref{pmnr:line:pick-alphas} of
Algorithm~\ref{alg:pmnr}, the optimizable parameter selection
heuristic \(\pickAlphas\) outputs optimizable parameters
\(\vecalphaselec, \vecalphagener, \vecalphafinal\) to be used in the
next stages of the PMNR paradigm. Then, at
line~\ref{pmnr:line:select-neurons}, \(\selectNeurons\) outputs a set
of activation neurons
\(\neurons = \{\hx{i_1}{j_1}, \ldots \hx{i_d}{j_d}\}\). Afterwards,
at line~\ref{pmnr:line:generate-pmnr}, \(\generatePMNR\) returns a
set of multi-neuron bounds \(\multiNeuronBounds\) which is a
polyhedron \(\sum_{k=1}^{d} ( \vecc{k} \hx{i_k}{j_k} - \vecC{k} \vecx{i_k} ) \le \vecdd\),
Finally, at line~\ref{pmnr:line:tighten}, \(\PostTighten\) returns
updated bounds \(\concreteBounds'\). For soundness, our framework
requires that \(\multiNeuronBounds\) is a lemma learned from
\(\concreteBounds \cup \singleNeuronBounds(\vecalphagener)\),
\(\concreteBounds'\) is a lemma learned from
\(\concreteBounds \cup \singleNeuronBounds(\vecalphafinal) \cup \multiNeuronBounds\),
and it holds true that \(\concreteBounds' \Ra \concreteBounds\)
(i.e. \(\concreteBounds'\) is stronger than \(\concreteBounds\)).

\begin{algorithm}[ht!]
	\small
	\begin{algorithmic}[1]
		\caption{\pmnr(\(N, \Din, \Dout)\)\label{alg:pmnr}}
		\WHILE {\(\neg \stopCondition\)()}
		\STATE {\(\concreteBounds, \singleNeuronBounds, \vecalphainiti \gets \singleNeuronTightening(N, \Din, \Dout)\)}\label{pmnr:line:single-neuron-tightening}
		\IF {\(\bot \in \concreteBounds\)}\label{pmnr:line:early-stopping}
		\STATE \textbf{break}
		\ENDIF
		\STATE {\(\vecalphaselec, \vecalphagener, \vecalphafinal \gets \pickAlphas(N, \concreteBounds, \singleNeuronBounds, \vecalphainiti)\)}\label{pmnr:line:pick-alphas}
		\STATE {\(\neurons \gets \selectNeurons(N, \concreteBounds, \singleNeuronBounds, \vecalphaselec)\)}\label{pmnr:line:select-neurons}
		\STATE {\(\multiNeuronBounds \gets \generatePMNR(N, \concreteBounds, \singleNeuronBounds, \vecalphagener, \neurons)\)}\label{pmnr:line:generate-pmnr}
		\STATE {\(\concreteBounds \gets \PostTighten(N, \Din, \Dout, \concreteBounds, \singleNeuronBounds, \vecalphafinal, \multiNeuronBounds)\)}\label{pmnr:line:tighten}
		\ENDWHILE
		\IF {\(\bot \in \concreteBounds\)}
		\STATE \textbf{break}
		\ENDIF
		\RETURN\(\concreteBounds\)
	\end{algorithmic}
\end{algorithm}

\mysubsection{Soundness.} It is straightforward to prove by induction
that the returned concrete bounds from \(\pmnr\) are sound and are
no less precise than those inferred by \(\singleNeuronTightening\) if
\(\pmnr\)'s soundness requirements hold. Formally:
\begin{theorem}\label{thm:soundness}
	If \(\pmnr\)'s requirements hold and \(\concreteBounds_{single},
	\concreteBounds_{\pmnr}\) are the concrete bounds yielded by
	\(\singleNeuronTightening\) and Algorithm~\ref{alg:pmnr}
	respectively, then \(Q \Ra \concreteBounds_{\pmnr}\) and
	\(\concreteBounds_{\pmnr} \Ra \concreteBounds_{single}\).
\end{theorem}

\section{Instantiating PMNR}\label{sec:instantiating-pmnr}

In Section~\ref{sec:pmnr-paradigm}, we described our approach for
performing partial multi-neuron tightening and proved soundness
properties. Here, we list specific heuristics we used to instantiate
the PMNR paradigm in our experiments. Our suggested heuristics build
on contemporary branching heuristics and bound tightening algorithms
that tolerate general non-linearities, and are therefore generally
applicable to multiple kinds of DNNs and activations.

\subsection{Neuron Selection}\label{instantiating-pmnr:subsec:neuron-selection}
\mysubsection{Neuron Selection with Symbolic Expressions (NSSE).}
The novel NSSE heuristic presented here, designed for neuron
selection from DNNs with arbitrary activation functions, is reworked
from the Bound Propagation with Shortcuts (BBPS) branching
heuristic~\cite{ShJiKoJaHsZh25}. BBPS, which supports arbitrary
non-linearities, estimates the lower bound on a DNN's output neuron
for all potential branchings and selects the neuron which is
projected to yield the maximal improvement. One major alteration
between BBPS and NSSE is that BBPS assigns a score for every pair of
neuron and possible branch, while NSSE assigns a score for every
unfixed-phase neuron.

\mysubsection{Calculating BBPS Scores.} To calculate the BBPS score
for the \supth{\(k\)} phase of neuron \(\x{i}{j}\), the single-neuron
relaxation-based tightening method is augmented to calculate linear
lower over-approximations on \(\vecx{L}\) in terms of \(\hx{i}{j}\),
as well as linear bounds on \(\hx{i}{j}\) in terms of \(\vecx{i}\),
which are sound when \(\x{i}{j}\) is in its \supth{\(k\)} phase. By
substituting the concrete bounds on \(\vecx{i}\) in these linear
over-approximations, a concrete linear bound for the output layer
is calculated. This lower bound is defined to be the BBPS score of
\supth{\(k\)} phase of neuron \(\x{i}{j}\), and it serves as a cheap
approximation on the post-branching lower bounds on the output. The
BBPS heuristic prioritizes neurons with the highest score, in an
attempt to prove the output domain \((0, \infty)\) is satisfiable.

\mysubsection{Calculating NSSE Scores.} To calculate the NSSE score
of a neuron \(\x{i}{j}\), we derive linear upper and lower bounds on
the output neuron \(\vecx{L}\) as a function of a chosen source
neuron \(\x{i}{j'}\) which are sound when \(\x{i}{j}\) is in its
\supth{\(k\)} phase, in a similar fashion to the calculation routine
of the BBPS heuristic. Then, we separately aggregate the linear upper
and lower over-approximations over all branches, producing two
symbolic expressions of \(\x{i}{j'}\). and the post-concretization
average range of these symbolic expressions is \(\x{i}{j}\)'s NSSE
score. Our heuristic picks the \(d\) highest-score unfixed-phase
neurons from the layer \(\ell\) with highest score-sum layer, unlike
BBPS which picks the highest-score neuron and branch.

\subsection{PMNR Generation}\label{instantiating-pmnr:subsec:pmnr-generation}
We will break down the novel \emph{Bounding hyper-planes via
Splitting and Optimization} (BHSO) paradigm for generating bounding
hyper-planes, described in Algorithm~\ref{alg:generate-pmnr}, and
apply it to PMNR generation. BHSO features elements from the
Branch-and-Bound paradigm~\cite{BuLuTuToKoKu20,ShJiKoJaHsZh25} and
preimage over-approximation~\cite{SuChZiKrHu23}, and applies them to
the problem of inferring bounding hyper-planes.

\mysubsection{Initial hyper-planes.}
Though BHSO applies to general bounding hyper-planes
\(\sum_{i=0}^{L-1} \vechC{i} \vechx{i} + \sum_{i=1}^{L} \vecC{i} \vecx{i} \le \vecdd\),
we focused in our evaluation on inferring multi-neuron bounds of the
form (similarly to~\cite{SiGaPuVe19}):

\[ \sum_{k \in [d]} \vecepsilon{k} \left(\hx{\ell}{j_k} - \vecWu{\ell}_{j_k:} \vecx{\ell} \right) \le \vecdd,\quad \sum_{k \in [d]} \vecepsilon{k} \left(\hx{\ell}{j_k} - \vecWl{\ell}_{j_k:} \vecx{\ell} \right) \le \vecdd.\]

To avoid re-optimizing \(\concreteBounds\) or \(\singleNeuronBounds\)
with \(\optimizePMNR\), we limited ourselves to vectors
\(\vecepsilonn \in \{-1, 0, 1\}^d\) with more than one non-zero
entry. The number of such vectors equals \(3^d - 2d - 1\), which
grows exponentially with \(d\): For \(d\) values of \(2, 3, 4\), the
quantity of bounding hyper-planes to be generated would be \(4\),
\(20\) and \(72\) respectively. We limit ourselves to
\(d \in \{2, 3\}\) selected neurons in order to ensure \(\pmnr\)
remains computationally affordable within our experiments.

\mysubsection{Optimizing hyper-planes.}
BHSO employs \(\optimizePMNR\) to refine the bias of all hyper-planes
defined above. In the case of ReLU networks, the INVPROP
algorithm~\cite{SuChZiKrHu23} might be used to instantiate it. To
support general networks, we employ a generalized version
of~\cite[Theorem 2, Appendix C]{SuChZiKrHu23}, which applies to
general activations \(\sig{i}\) and input domains \(\Din\) and allows
to optimize current hyper-planes depending on previous ones. See
Appendix~\ref{appendix:theorem-two-details} for details on the
generalized theorem, and Appendix~\ref{appendix:theorem-proof} for
proof. We utilize this theorem to optimize multi-neuron bounds
sequentially given \(\concreteBounds, \singleNeuronBounds\) and
previously optimized multi-neuron bounds \(\multiNeuronBounds\) with
PGD.

\mysubsection{Optimizing Further with General Branching.}
BHSO incorporates branching in order to learn more precise bounds
from \(\optimizePMNR\). Like NSSE, it assumes all chosen neurons
could be partitioned into several branches, for instance, via ReLU
splitting or GenBaB~\cite{ShJiKoJaHsZh25}. \(\optimizePMNR\) operates
several times per hyper-plane, with each run superseding a
pre-activation value's bounds with those of the current branch
combinations, and substituting the neurons' linear
over-approximations with the corresponding, more accurate ones. The
hyper-plane's new bias \(\vecd{k}\) is the weakest bound among all
branch combinations, thereby ensuring the soundness of BHSO is not
impaired by the existence of infeasible branch combinations (for
which the dual problem solved by INVPROP or
Theorem~\ref{thm:generalized-invprop} is unbounded).

As an illustration, here are the main steps that the BHSO paradigm
might perform to deduce a hyper-plane involving the \(\relu\) neurons
\(\hx{2}{0}, \hx{2}{1}\) from the example network in
Fig.~\ref{fig:example-network}. An initial hyper-plane would be
directly derived via \(\optimizePMNR\), leveraging the neurons'
unfixed-phase linear bounds. To tighten it further, \(\optimizePMNR\)
will be called four additional times, per each combination of the
neurons being in their active
\(\x{2}{0} \le \hx{2}{0} \le \x{2}{0},\enspace\x{2}{1} \le \hx{2}{1} \le \x{2}{0}\)
or inactive phase
\(0 \le \hx{2}{0} \le 0,\enspace0 \le \hx{2}{1} \le 0\), while
superseding the neurons' unfixed-phase linear bounds with precise
per-branch bounds. Ultimately, BHSO will choose the loosest bound
found.

\mysubsection{Infeasible Branches Detection.}
It is possible to identify some branch combinations which are
infeasible by calculating an upper bound for the bounding
hyper-planes (e.g. with simple concretization) and comparing it to
the bias resulted by invoking \(\optimizePMNR\). These findings might
be integrated with the next steps of the larger Branch-and-Bound
paradigm, though we have not explored this direction yet.

\begin{algorithm}[htb!]
	\small
	\begin{algorithmic}[1]
		\caption{\generatePMNR(\(N, \concreteBounds, \singleNeuronBounds, \vecalphagener, \neurons)\)\label{alg:generate-pmnr}}
		\STATE {\(\branchPoints,\branchBounds \gets \Branches(N, \concreteBounds, \singleNeuronBounds, \vecalphaselec)\)}
		\STATE {\(\infeasibleBranches \gets \emptyset\)}
		\STATE {\(\multiNeuronBounds \gets \emptyset\)}
		\STATE {\(m,\vecC{i}, \vechC{i}, \vecd{i}, \vecdfeasi{i} \gets \)}\\
		\(\initialPMNR(N, \concreteBounds, \singleNeuronBounds, \vecalphagener, \vecC{i}, \vechC{i})\)
		\FOR {\(k \in 1, \ldots m\)}
		\STATE {\(\vecdoptim{k} \gets \)}\\
		\(\optimizePMNR(N, \concreteBounds, \singleNeuronBounds, \vecalphagener, \vecC{i}, \vechC{i}, \vecd{i}, \multiNeuronBounds)\)
		\STATE {\(\vecdbranc{k} \gets -\infty\)}
		\FOR {\(\branchCombination \text{ in } \branchCombinations(\neurons, \branchBounds)\)}
		\STATE{\(\vecdbranc{k} \gets\)}\\
		\(\optimizePMNR(N, \concreteBounds, \singleNeuronBounds, \vecalphagener, \vecC{i}, \vechC{i}, \vecd{i}, \multiNeuronBounds)\)
		\IF {\(\vecdbranc{k} > \vecdfeasi{k}\)}
		\STATE{\(\infeasibleBranches \gets \infeasibleBranches \cup \{\branchCombination\}\)}
		\ENDIF
		\STATE{\(\vecdoptim{k} \gets \fmin{\vecdoptim{k}, \vecdbranc{k}}\)}
		\ENDFOR
		\STATE{\(\vecd{k} \gets \fmax{\vecd{k}, \vecdoptim{k}}\)}
		\STATE{\(\multiNeuronBounds \gets \multiNeuronBounds \cup \{ \sum_{i=0}^{L-1} \vechC{i}_{:k} \vechx{i} + \sum_{i=1}^{L} \vecC{i}_{:k} \vecx{i} \le \vecd{k}\}\)}
		\ENDFOR
		\RETURN\(\multiNeuronBounds\)
	\end{algorithmic}
\end{algorithm}

\subsection{Other Heuristics}\label{instantiating-pmnr:subsec:other-heuristics}
In this subsection we describe other heuristics employed in our
evaluation.

\mysubsection{Single-Neuron Tightening and Stopping Criteria.}
We instantiated the method \(\singleNeuronTightening\) with
DeepPoly~\cite{GaGePuVe19}, which rapidly gathers concrete bounds by
propagating single-neuron bounds across a DNN via concretization and
back-substitution. If the \(\bot\) becomes an element of
\(\concreteBounds\) during at any point during
Algorithm~\ref{alg:pmnr}, then \(\pmnr\) terminates as the
verification query \(Q\) is proven to be unsatisfiable. The stopping
criterion \(\stopCondition\), which controls the execution of the
main loop of \(\pmnr\), holds when any of these conditions is met:
(i) the main loop of Algorithm~\ref{alg:pmnr} has completed \(n\)
iterations, where \(n\) is a user-defined budget parameter; or (ii)
the operation of \(\PostTighten\) at line~\ref{pmnr:line:tighten} of
Algorithm~\ref{alg:pmnr} has not resulted in any revision to
\(\concreteBounds\).

\mysubsection{Optimizable Parameters.}
Among the algorithm four tunable parameters, the first three
\(\vecalphainiti = \vecalphaselec = \vecalphagener\) are chosen as
detailed at~\cite{WaPiWhYaJa18}, whilst \(\vecalphafinal\) is defined
as such: Linear over-approximations of \(N\)'s output layer in terms
of its input layer are produced through Symbolic
Intervals~\cite{WaPeWhYaJa18}, following which local
optimization~\cite[Section 4.2, Appendix F.]{ZhWaKw24} is applied to
select \(\vecalphafinal\) which minimizes the volume of the
input-space polytope created by them.

\mysubsection{Final Tightening.}
To further tighten \(\concreteBounds\) given multi-neuron bounds,
\(\PostTighten\) capitalizes on a modified version of the LP-based
Forward- Backward Abstract Interpretation~\cite{WuBaShNaSi22}
framework. It consists of a forward pass, during which only the
subset of linear bounds from
\(\Din \cup \Dout \cup \concreteBounds \cup \singleNeuronBounds \cup \multiNeuronBounds\)
containing neurons from current or preceding layers are counted among
the constraints of the LPs solved, as well as a backward pass, in
which only those containing neurons from current or subsequent layers
are included.

\subsection{Heuristics For PMNR-ALL}\label{instantiating-pmnr:subsec:heuristics-for-pmnr-all}
For the purpose of fairly comparing PMNR instantiated with the
heuristics described in previous subsections to existing Multi Neuron
Relaxation approaches, we introduce another instantiation of the PMNR
paradigm called \(\pmnrall\).

It generates multi-neuron bounds for nearly all activation neurons
from all layers using the same heuristics in
Subsections~\ref{instantiating-pmnr:subsec:pmnr-generation}
and~\ref{instantiating-pmnr:subsec:other-heuristics}, although it
differs from our main instantiation of \(\pmnr\) in regard to neuron
selection. While \(\pmnr\) chooses \(d\) neurons from a single layer
with the NSSE heuristic, \(\pmnrall\) selects a set containing all
groups of \(d\) consecutive unfixed-phase activation neurons from all
layers. Following neuron selection, both instantiations produce
hyper-planes involving each group of \(d\) neurons separately, via
BHSO.

\(\pmnr\) constitutes a middle-ground between
\(\singleNeuronTightening\) and \(\pmnrall\) in regard to performance
and tightness. \(\pmnr\) improves on \(\singleNeuronTightening\) by
producing hyper-planes only involving \(d\) heuristically selected
neurons, whereas \(\pmnrall\) does so by by generating multi-neuron
bounds for nearly all neurons. Notably, in \(\pmnrall\) the number of
hyper-planes to be calculated scales linearly in the size of the DNN,
while in \(\pmnr\) it would be constant. Combined with the fact that
the computational resources necessary to generate a single
hyper-plane also grows with the DNN's size, it follows that the
bounds discovered by \(\pmnrall\) are likely to be stronger than
these deduced by \(\pmnr\), though they are more computationally
expensive to obtain.

\subsection{Running Example}\label{instantiating-pmnr:subsec:running-example}
Here is a demonstration of \(\pmnr\) on an example query \(Q\),
featuring the network depicted in Fig.~\ref{fig:example-network}.

\mysubsection{DeepPoly Fails to Verify \(Q\).} Consider once more the
neural network \(N\) from Fig.~\ref{fig:example-network}, and domains
\(\Din=[-1, 1]^2\) and \(\Dout=(-\infty, 0)\). Executing DeepPoly (as
the instantiation of \(\singleNeuronTightening\)) yields the bounds
depicted in Appendix~\ref{appendix:running-example-details}. DeepPoly
does not manage to prove that the verification query
\(Q = (N, \Din, \Dout)\) is \(\UNSAT\), because the computed output
layer's bounds \(\x{3}{0} \in [-0.15, 40.1]\) are not adequately
strong. Thus, we progress to the ensuing stages of the PMNR paradigm.

\mysubsection{Running Example Heuristics.} For the running example,
we employed simplified tunable parameters and neuron selection
heuristics to demonstrate \(\pmnr\). First,
\(\frelu{\x{i}{j}} \ge \x{i}{j}\) and \(\fabs{\x{i}{j}} \ge 0\) are
the linear lower bounds of our choice for unfixed-phase ReLU neurons
and Abs neurons, respectively. Moreover, rather than computing NSSE
scores for all unfixed-phase activation neurons, we use the span of
each neuron's concrete bounds as its score: i.e.,
\(\score{i}{j} = \upperb{i}{j} - \lowerb{i}{j}\) is the score of
neuron \(\hx{i}{j}\).

\mysubsection{Multi-Neuron Relaxation Solves \(Q\).} We first
demonstrate how \(Q\) is solved with \(\pmnrall\), a Multi-Neuron
Relaxation-based approach; and then proceed to show that PMNR
likewise solves it while requiring less computational effort.

PMNR-ALL selects all unfixed-phase neurons in \(N\):
\(\hx{1}{1}, \hx{1}{2}, \hx{2}{0}, \hx{2}{1}\). It applies
multi-neuron bound generation with BHSO, and, given the linear
over-approximations discovered by DeepPoly, it produces \(12\)
multi-neuron bounds of the form
\(\vecepsilon{1} \hx{1}{0} + \vecepsilon{2} \hx{1}{1} \le t\),
\(\vecepsilon{1} (\hx{2}{0} - \x{2}{0}) + \vecepsilon{2} (\hx{2}{1} -
\x{2}{1}) \le t\) and
\(\vecepsilon{1} (\hx{2}{0} - \tfrac{7}{8}\x{2}{0}) + \vecepsilon{2}
(\hx{2}{1} - \tfrac{7}{12}\x{2}{1}) \le t\) for
\(\vecepsilonn \in \{-1, 1\}^2\)
(see Appendix~\ref{appendix:running-example-details} for the
resulting hyper-planes).

Applying \(\PostTighten\) results in these concrete bounds:
\(\lowerhb{2}{0} \gets 0,\)\, \(\lowerhb{2}{1} \gets 0\),
\(\lowerhb{3}{0} \gets 0.1\), \(\upperhb{3}{0} \gets 26.1\).
Augmenting DeepPoly with partial multi-neuron relaxation yielded the
tightened output layer's bounds \(\x{3}{0} \in [0.1, 26.1]\), which,
when intersected with the output domain \(\Dout = (-\infty, 0)\),
results in empty concrete bounds \(Q \Ra \bot\). The stronger
relaxations learned by PMNR-ALL thus allowed proving that \(Q\) is
\(\UNSAT\).

\sloppy
\mysubsection{PMNR Solves \(Q\) more Quickly.} The unfixed-phase
neurons of \(N\) are \(\hx{1}{1}, \hx{1}{2}, \hx{2}{0}, \hx{2}{1}\),
with associated scores  \(\score{1}{1}=10\), \(\score{1}{2}=2\),
\(\score{2}{0}=8\), \(\score{2}{1}=12\). Because \(\vechx{2}\) is the
layer with greatest neuron score sum, the \(d=2\) highest-score
neurons from it (namely, \(\hx{2}{0}\) and \(\hx{2}{1}\)) are chosen
per our BHSO paradigm. PMNR produces eight multi-neuron bounds of the
form
\(\vecepsilon{1} (\hx{2}{0} - \x{2}{0}) + \vecepsilon{2} (\hx{2}{1} - \x{2}{1}) \le t, \vecepsilon{1} (\hx{2}{0} - \tfrac{7}{8}\x{2}{0}) + \vecepsilon{2} (\hx{2}{1} - \tfrac{7}{12}\x{2}{1}) \le t\)
for \(\vecepsilonn \in \{-1, 1\}^2\), which are listed in
Appendix~\ref{appendix:running-example-details}.

Applying \(\PostTighten\) yields the same bounds computed by
PMNR-ALL, proving \(Q\) is \(\UNSAT\). Thus, despite having
calculated fewer hyper-planes via BHSO than PMNR-ALL, PMNR manages to
solve \(Q\). The ratio between the number of hyper-planes produced by
the Multi-Neuron Relaxation-based PMNR-ALL versus PMNR scales
linearly with \(N\)'s size, implying that PMNR's advantage over
PMNR-ALL should become even clearer for larger DNNs.

\section{Experiments and Evaluation}\label{sec:experiments-and-evaluation}

\subsection{Implementation}\label{experiments-and-evaluation:subsec:implementation}
For our evaluation, we implemented the PMNR paradigm with the
heuristics defined in
Subsections~\ref{instantiating-pmnr:subsec:neuron-selection}--\ref{instantiating-pmnr:subsec:other-heuristics}
within the SMT-based Marabou verification
tool~\cite{KaHuIbJuLaLiShThWuZeDiKoBa19}, and compared it to the
DeepPoly~\cite{GaGePuVe19} Symbolic Bound Tightening framework. In
addition, we implemented PMNR-ALL defined in
Subsection~\ref{instantiating-pmnr:subsec:heuristics-for-pmnr-all} as
a representative multi-neuron relaxation method, and compared it to
our approach. See Appendix~\ref{appendix:evaluation-marabou} for a
more detailed description of our findings.

We evaluated our approach on \(\ell_\infty\)-local robustness queries
with fully-connected FFNNs, trained on the MNIST
dataset~\cite{LeBeHa98}, including both piecewise-linear (PL) and
non-piecewise-linear (NPL) activations. Local robustness verification
pertains to the robustness of a DNN to small perturbations around a
given input. Formally, given a DNN \(N\), an input \(x_0\) and
positive reals \(\varepsilon, \delta\), \(\ell_\infty\)-local
robustness queries have
\(\Din = \{x : \| x - x_0 \|_\infty \le \varepsilon\}\) and
\(\Dout = \{x : \| N(x) - N(x_0) \|_\infty \le \delta\}\).

Marabou's SMT Solver is interleaved with calls to bound tightening
procedures (by default, DeepPoly)~\cite{MarabouII2025}. In our
implementation, the initial bound tightening algorithm is replaced by
our implementation of PMNR or PMNR-ALL, while DeepPoly is called for
subsequent bound tightening. We have opted to employ DeepPoly for
subsequent tightening since invoking PMNR repeatedly would incur an
intolerable computational toll. This experimental setup guarantees
the only difference between base Marabou, PMNR-enhanced and
PMNR-ALL-enhanced Marabou results from augmenting the first run of
DeepPoly with PMNR or PMNR-ALL, with the aim of discovering stronger
bounds. We evaluated Marabou with DeepPoly, \(\pmnr\) and
\(\pmnrall\) as the initial bound tightening method on \(344\) local
robustness queries overall, including \(272\) queries for PL networks
and \(72\) queries for NPL networks. The architectures of all the
FFNNs in our evaluation are specified in
Appendix~\ref{appendix:experimental-details}.

The external LP solver Gurobi~\cite{Gurobi} served to assist the PMNR
paradigm's final bound tightening method \(\PostTighten\). For our
experiments in this section, we set \(n=10\) for the stopping
condition for PMNR and PMNR-ALL. All experiments were conducted on
dual-core machines with 4GB of memory, running Debian 12, with a
timeout of \(T=14100\) seconds (235 minutes).

\subsection{FFNNs with Piecewise-Linear Activations}\label{experiments-and-evaluation:subsec:piecewise-linear-ffns}
For our benchmarks on DNNs with piecewise-linear activations, we
experimented on fully-connected networks featuring the \(\relu\),
\(\leakyrelu\), \(\maxxx\) (Max Pool)~\cite{GaGePuVe19} and
\(\sign\)~\cite{AmWuBaKa21} activations.

We trained the \textsc{ReluSignMax} and \textsc{LeakyRelu}
\(14\times28\) FFNNs ourselves using the PyTorch
library~\cite{PyTorch}, and verified local robustness for the first
image in the MNIST test set with arbitrarily selected \(\varepsilon\)
values of \(0.02, 0.04, 0.06, 0.08\), thereby producing \(36\)
queries per DNN. As for the remaining two FFNNs, we used the two
benchmarks denoted as \(\MNIST{1}, \MNIST{2}\)
in~\cite{WuBaShNaSi22}, which consist of verifying local robustness
around the first 100 MNIST test images with \(\varepsilon=0.02\).

The results of our experiments on PL networks is summarized at
Table~\ref{tab:results-piecewise-linear}. The ``Verified'' column
represents the number of robustness queries which Marabou verified as
either satisfiable or unsatisfiable within the time frame of
\(14100\) seconds, and the ``Time'' column represents the average
time required to verify a query (in seconds), computed over the
queries that the solver successfully verified.

These results highlight the superiority of our approach over both
Single-Neuron Relaxation and Multi-Neuron Relaxation: thanks to the
tighter hyper-planes discovered by PMNR, PMNR-enhanced Marabou has
verified \(245\) queries out of \(272\), a \(40\%\) improvement over
base Marabou which verified only \(156\) queries, and a \(612\%\)
improvement over PMNR-ALL which verified merely \(40\) queries within
the time limit due to its higher runtime overhead.

\begin{table*}[htb]
    \centering
	\caption{Comparing \(\pmnr\) to DeepPoly and \(\pmnrall\) on piecewise-linear networks.}
	\begin{tabular}{|c|c|cc|cc|cc|}
		\hline
		\multirow{2}{*}{Model}         & \multirow{2}{*}{Queries}      &
		\multicolumn{2}{c|}{\(\pmnr\)} & \multicolumn{2}{c|}{DeepPoly} & \multicolumn{2}{c|}{\(\pmnrall\)}                                                                                                                                                                \\ \cline{3-8}
		                               &                               & \multicolumn{1}{c|}{Solved \(\uparrow\)} & Time \(\downarrow\) & \multicolumn{1}{c|}{Solved \(\uparrow\)} & Time \(\downarrow\) & \multicolumn{1}{c|}{Solved \(\uparrow\)} & Time \(\downarrow\) \\ \hline
		\LeakyReluFive                 & 100                           & \multicolumn{1}{c|}{\textbf{100}}        & \textbf{67}         & \multicolumn{1}{c|}{97}                  & 129                 & \multicolumn{1}{c|}{40}                  & 5003                \\ \hline
		\LeakyReluEight                & 100                           & \multicolumn{1}{c|}{\textbf{98}}         & \textbf{294}        & \multicolumn{1}{c|}{33}                  & 501                 & \multicolumn{1}{c|}{0}                   & \(\infty\)          \\ \hline
		\LeakyReluFourteen             & 36                            & \multicolumn{1}{c|}{\textbf{29}}         & 1321                & \multicolumn{1}{c|}{26}                  & \textbf{16}         & \multicolumn{1}{c|}{0}                   & \(\infty\)          \\ \hline
		\ReluSignMax                   & 36                            & \multicolumn{1}{c|}{\textbf{18}}         & \textbf{6138}       & \multicolumn{1}{c|}{\textbf{18}}         & 6742                & \multicolumn{1}{c|}{0}                   & \(\infty\)          \\ \hline
		Total                          & 272                           & \multicolumn{1}{c|}{\textbf{245}}        & \textbf{752}        & \multicolumn{1}{c|}{174}                 & 866                 & \multicolumn{1}{c|}{40}                  & 5003                \\ \hline
	\end{tabular}
	\label{tab:results-piecewise-linear}
\end{table*}

\subsection{FFNNs with Non-Piecewise-Linear Activations }\label{experiments-and-evaluation:subsec:non-piecewise-linear-ffns}
For evaluation on non-piecewise-linear activations, we used the
PyTorch library to train MNIST classifiers containing the
\(\sigmoid\)~\cite{GaGePuVe19}, \(\bilinear\)
(Multiplication)~\cite{ShJiKoJaHsZh25} and
\(\softmax\)~\cite{WeWuWuChBaFa23} activations.
We executed Marabou on \(36\) local robustness queries per network,
characterized in the same way as the \textsc{LeakyRelu}
\(14\times28\) benchmarks. Summary statistics for our experiments on
NPL networks are visible at Table~\ref{tab:results-npl}. Notably,
PMNR-enhanced Marabou verified \(56\) queries out of \(72\), an
\(100\%\) increase over base Marabou (\(28\) verified) and a \(3\%\)
increase over PMNR-ALL-enhanced Marabou (\(54\) verified). Overall,
out of the \(344\) robustness queries tested, PMNR-enhanced Marabou
successfully solved \(301\) queries, an \(49\%\) improvement over
base Marabou which solved 202 queries; and a \(220\%\) improvement
over PMNR-ALL-enhanced Marabou which solved 94. Employing PMNR within
Marabou resulted in an average time requirement for verification of
619 seconds, which is \(17\%\) faster than DeepPoly (751 seconds) and
\(86\%\) faster than PMNR-ALL (4622 seconds).

\begin{table*}[htb]
	\centering
	\caption{Comparing \(\pmnr\) to DeepPoly and \(\pmnrall\) on NPL FFNNs.}
	\begin{tabular}{|c|c|cc|cc|cc|}
		\hline
		\multirow{2}{*}{Model}         & \multirow{2}{*}{Queries}       &
		\multicolumn{2}{c|}{\(\pmnr\)} & \multicolumn{2}{c|}{DeepPoly}  & \multicolumn{2}{c|}{\(\pmnrall\)}                                                                                                                                                                \\ \cline{3-8}
		                               &                                & \multicolumn{1}{c|}{Solved \(\uparrow\)} & Time \(\downarrow\) & \multicolumn{1}{c|}{Solved \(\uparrow\)} & Time \(\downarrow\) & \multicolumn{1}{c|}{Solved \(\uparrow\)} & Time \(\downarrow\) \\ \hline
		\ReluBilinearSoftmax           & 36                             & \multicolumn{1}{c|}{29}                  & 54                  & \multicolumn{1}{c|}{10}                  & \textbf{46}         & \multicolumn{1}{c|}{\textbf{36}}         & 6485                \\ \hline
		\LeakyReluSigmoid              & 36                             & \multicolumn{1}{c|}{\textbf{27}}         & 22                  & \multicolumn{1}{c|}{18}                  & \textbf{19}         & \multicolumn{1}{c|}{18}                  & 50                  \\ \hline
		Total                          & 72                             & \multicolumn{1}{c|}{\textbf{56}}         & 38                  & \multicolumn{1}{c|}{28}                  & \textbf{28}         & \multicolumn{1}{c|}{54}                  & 4340                \\ \hline
	\end{tabular}
	\label{tab:results-npl}
\end{table*}

\begin{figure}[htb]
	\includegraphics[width=\linewidth,trim=0ex 0ex 0ex 0.5ex,clip]{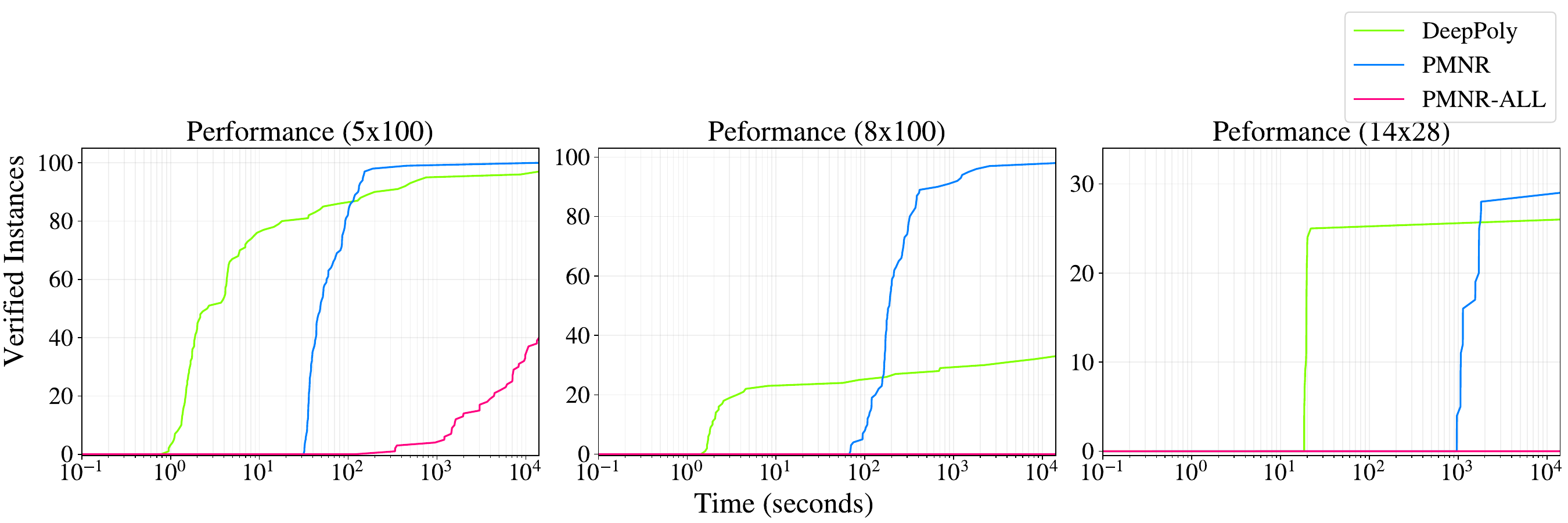}
	\includegraphics[width=\linewidth,trim=0ex 0ex 0ex 0.5ex,clip]{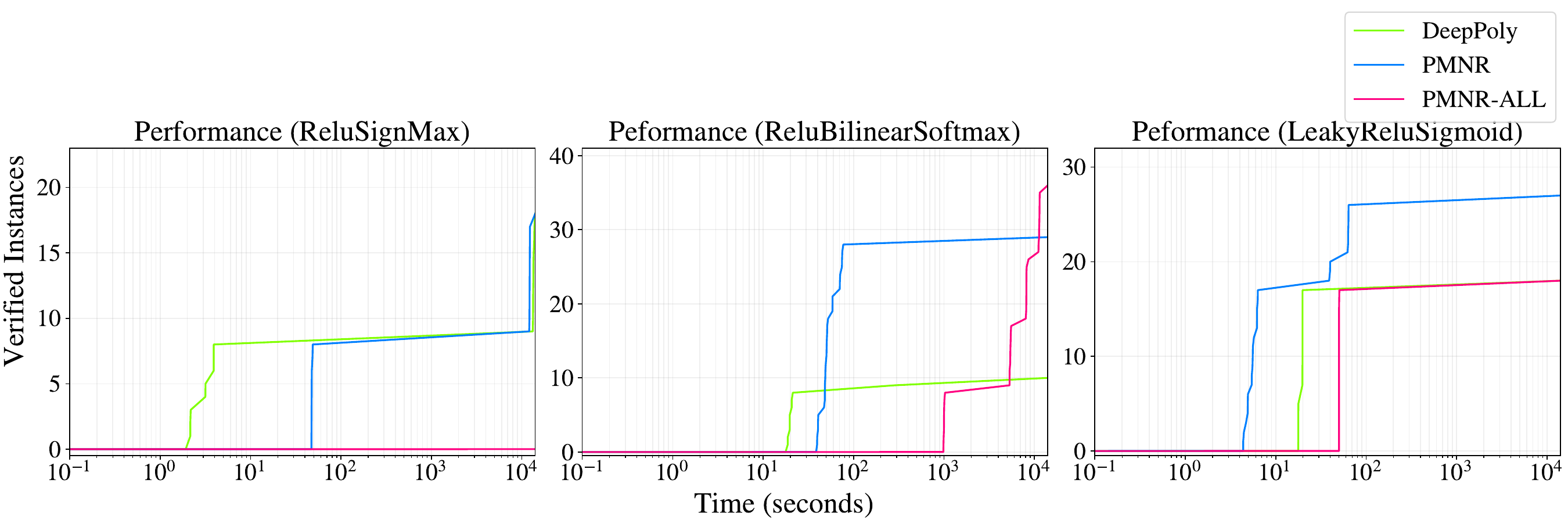}
	\caption{Cumulative instances verified vs.\ time (seconds,
	         log scale) for LeakyReLU FFNNs (Top) and other
			 activations FFNNs (bottom).}\label{fig:cactus-plot}
\end{figure}

\begin{figure}[htb]
	\includegraphics[width=0.48\linewidth,trim=0ex 0ex 0ex 0.5ex,clip]{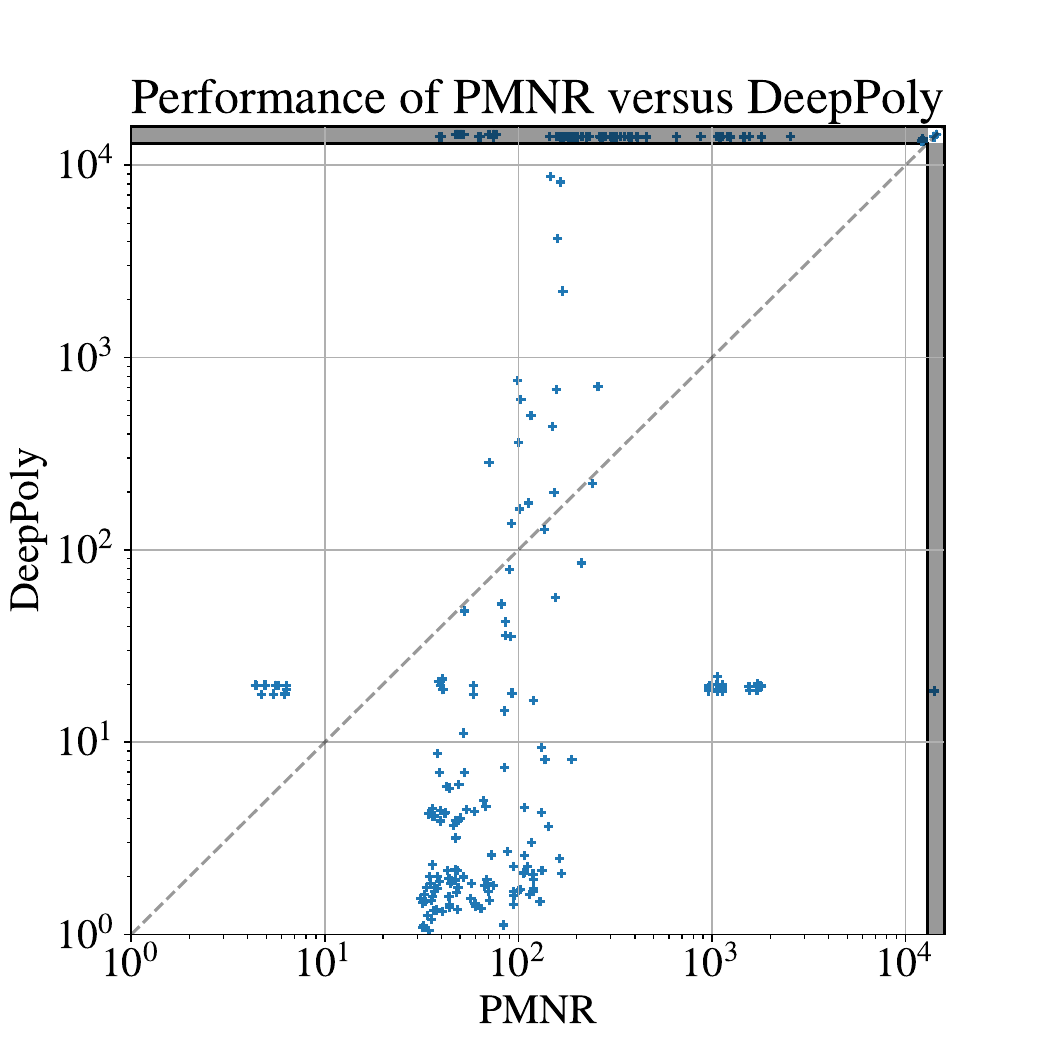}
	\includegraphics[width=0.48\linewidth,trim=0ex 0ex 0ex 0.5ex,clip]{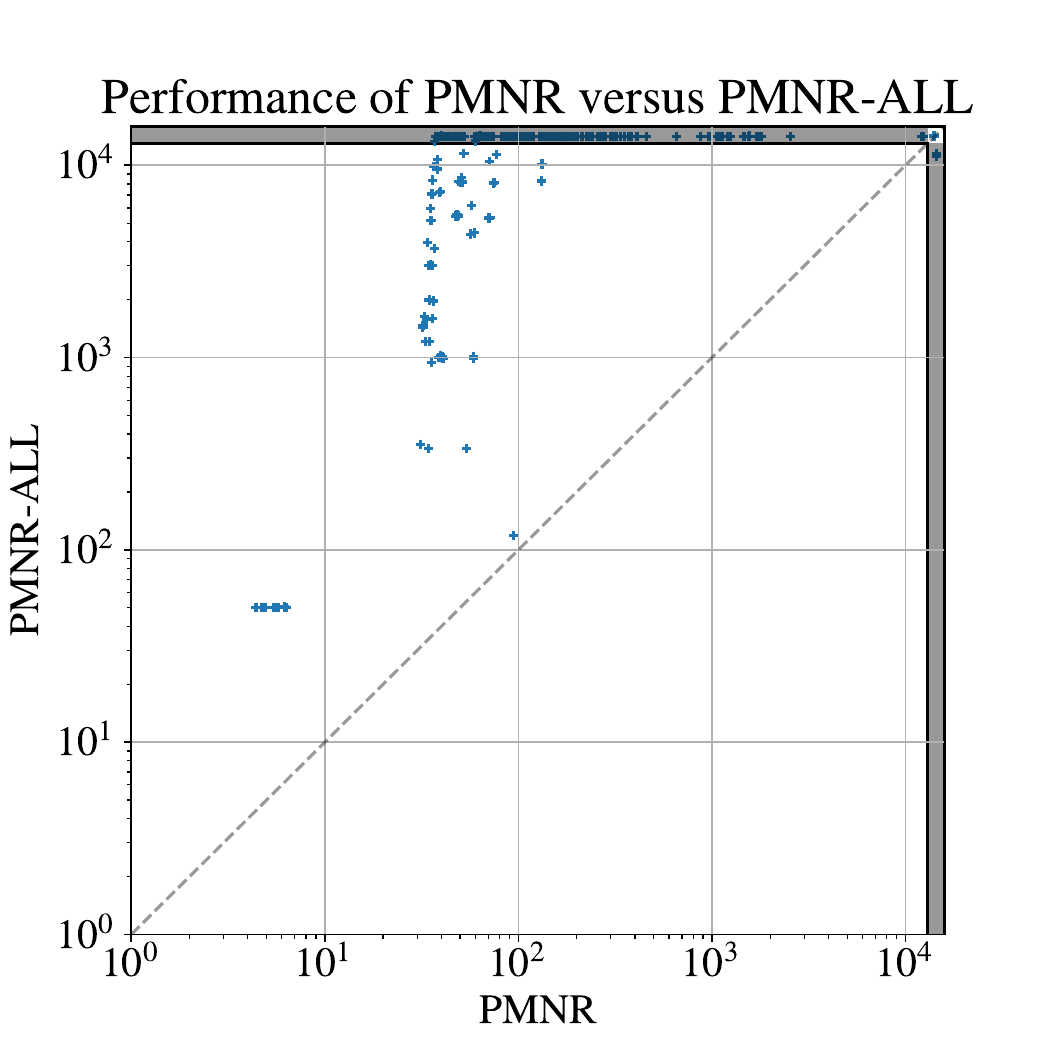}
	\caption{Comparing the runtimes (seconds, logscale) of
	         PMNR-enhanced Marabou to base Marabou (Left) and to
			 PMNR-ALL-enhanced Marabou (right).}\label{fig:scatter-plot}
\end{figure}

Fig.~\ref{fig:scatter-plot} compares the runtime of PMNR-enhanced
Marabou to base Marabou and PMNR-ALL-enhanced Marabou for both
classes of benchmarks, while Fig.~\ref{fig:cactus-plot} displays the
cumulative runtime of Marabou for every model. Both figures show
that, for all models beside \(\ReluBilinearSoftmax\), PMNR-enhanced
Marabou solved all queries more rapidly compared to PMNR-ALL-enhanced
Marabou --- due to the latter's higher associated computational
complexity. Further, the figures show that  for all models beside
\(\LeakyReluSigmoid\) there exist several  dozens of easier queries,
which are solved quickly by both methods and for which base Marabou
runs faster than PMNR-enhanced Marabou --- since the tighter concrete
bounds discovered by augmenting DeepPoly with PMNR cause an
unnecessary overhead. Nonetheless, the tighter bounds revealed by
PMNR assist Marabou in solving the remaining instances that base
Marabou would take longer to verify, or fail to verify within the
timeout. This is readily apparent in Fig.~\ref{fig:cactus-plot}, as
the graph of the cumulative number of instances PMNR-enhanced Marabou
solved ``catches up'' to the graph of base Marabou and surpasses it.


\section{Related Work}\label{sec:related-work}

\mysubsection{Bound Tightening for DNN Verification.}
The problem of DNN verification has been thoroughly studied in recent
years, bringing about various approaches to tackle this problem,
including BaB-based~\cite{BuLuTuToKoKu20,WaZhXuLiJaHsKo21,FeNuJoVe22}
and SMT-based
techniques~\cite{KaBaDiJuKo17,KaHuIbJuLaLiShThWuZeDiKoBa19,MarabouII2025},
abstraction-refinement~\cite{ElGoKa20}, techniques featuring
LP~\cite{Eh17}, MILP solvers~\cite{TjXiTe17,ZeWuBaKa22}, Lipschitz
bounds~\cite{SzZaSuBrErGoFe13} and other approaches~\cite{LiXiLi23}.

Our work focuses on bound tightening, which is a key element in many
DNN verification techniques. Since symbolic bound tightening methods
were first introduced~\cite{WaPeWhYaJa18,WaPiWhYaJa18,GaGePuVe19},
various techniques have been devised to improve upon them --- for
instance, by forward-backward abstract
interpretation~\cite{WuBaShNaSi22}, derivation of over-approximations
on the inputs with dual optimization~\cite{SuChZiKrHu23} or with
optimizable parameters~\cite{XuZhWaWaJaLiHs20,ZhWaKw24}, reducing
errors in symbolic bound propagation~\cite{ZeWuBaKa22}, spurious
region-guided refinement~\cite{YaLiLiHuWaSuXuZh21}, inferring
multi-neuron bounds from MILP coefficients~\cite{ZhWaLiLiJaHsKo22} or
from prior branching performed~\cite{ZhBrHaZh24}, and producing
multi-neuron bounds for all neurons~\cite{SiGaPuVe19,MuMaSiPuVe22}.

Building on this large body of existing work, our Partial
Multi-Neuron Relaxation (PMNR) approach heuristically selects neurons
and generates multi-neuron constraints only involving them and their
source neurons. Our framework is general, and is novel in the sense
that it allows selecting neurons arbitrarily, without needing to
perform actual branching, calculating MILP coefficients, or
generating multi-neuron bounds for all neurons.

\mysubsection{Dual Optimization of Multi-Neuron Bounds.} The INVPROP
algorithm~\cite{SuChZiKrHu23} receives concrete bounds and
single-neuron bounds for a ReLU DNN, and uses them to learn bounding
hyper-planes using dual linear optimization. INVPROP can be used for
generating multi-neuron bounds for ReLU networks, and it inspired our
generalized dual optimization method (see
Appendices~\ref{appendix:theorem-two-details}
and~\ref{appendix:theorem-proof}) --- which played a key part in our
heuristic for generating multi-neuron bounds for general activations,
as detailed in
Subsection~\ref{instantiating-pmnr:subsec:pmnr-generation}. We
further integrated the GenBaB branch-and-bound method for arbitrary
non-linearities~\cite{ShJiKoJaHsZh25} in our BHSO framework to
further optimize hyper-planes. Finally, the dual optimization
technique \(\beta\)-CROWN~\cite{WaZhXuLiJaHsKo21}, which encodes
split-neuron constraints using optimizable parameters, has been
successfully applied to obtain multi-neuron bounds for ReLU
networks~\cite{ZhWaLiLiJaHsKo22}. Integrating this technique with Generalized
INVPROP while supporting arbitrary activation functions remains an
intriguing direction for future work.

\mysubsection{Other Verification-Related Tasks.} The Marabou verifier
and other verification tools have been successfully applied to a wide
array of tasks, including verification of
binarized~\cite{AmWuBaKa21}, quantized~\cite{HuWuYaDaWuZhBa24} and
recurrent~\cite{JaBaKa20} neural networks, verification of aerospace
controllers~\cite{MaAmWuDaNeRaMeDuHoGaShKaBa24} and
reinforcement-learning systems~\cite{MaAmWuDaNeRaMeDuGaShKaBa24},
proof production and
minimization~\cite{IsBaZhKa22,ElIsKaLaWu25,IsReWuBaKa26}, ensemble
selection~\cite{AmZeKaSc22} and multi-layer
modification~\cite{ReKa22}. Our proposed approach for extending bound
tightening could benefit these tasks.

\section{Conclusion and Future Work}\label{sec:conclusion-and-future-work}
We presented our approach to augment existing Single Neuron
Relaxation-based bound tightening methods by learning multi-neuron
bounds featuring only a heuristically selected subset of all neurons.
We achieved this by designing new heuristics for neuron selection
(NSSE) and multi-neuron bound generation by altering contemporary
branching heuristics and bound tightening algorithms. Implementing
PMNR in Marabou has successfully resulted in the derivation of
tighter bounds at the cost of a runtime overhead for simpler queries,
and we seek to explore directions to improve it further. Some of our
directions for future work include: Evaluating our approach over
non-MNIST benchmarks and comparing it to, or combining it with, other related approaches, including PRIMA~\cite{MuMaSiPuVe22},
k-ReLU~\cite{SiGaPuVe19} or \(\beta\)-CROWN~\cite{WaZhXuLiJaHsKo21};
adding support for GPU-based parallelization; improving our NSSE and
BHSO methods, by integrating the resulting multi-neuron bounds and
BHSO-inferred infeasible branch combinations with search-based
techniques; combining our approach with automatic inferring of
single-neuron linear over-approximations~\cite{PaWa22} to support
arbitrary black-box activations, without expert-designed linear
bounds.

\mysubsection{Acknowledgments.}\label{sec:acknowledgments}
This research was partially supported by a grant
from the Israeli Science Foundation (grant number 558/24). In
addition, this work was partially funded by the European Union
(ERC, VeriDeL, 101112713). Views and opinions expressed are however
those of the authors only and do not necessarily reflect those of
the European Union or the European Research Council Executive Agency.
Neither the European Union nor the granting authority can be held
responsible for them.

\bibliographystyle{abbrv}
\bibliography{bibliography}

\newpage
\noindent
{\huge{Appendix}}

\appendix

\section{Experimental Details}\label{appendix:experimental-details}

\mysubsection{Network Architectures.} The architectures of all six
MNIST classifiers which were included in our local robustness
benchmarks throughout this paper are shown in
Table~\ref{tab:network-topologies}. The last two neural networks,
\(\ReluBilinearSoftmax\) and \(\LeakyReluSigmoid\), feature
non-piecewise-linear activations; whereas the first four consist
entirely of piecewise-linear ones.

\begin{table}[htb]
    \centering
	\caption{The FFNNs used in our experiments.}
	\begin{tabular}{|l|l|l|l|l|}
		\hline
		Dataset                & Model                & \(\begin{matrix}\text{Hidden}\phantom{\,}\\\text{Neurons}\end{matrix}\) & \(\begin{matrix}\text{Hidden}\\\text{Layers}\end{matrix}\) & Activations                                                                            \\ \hline
		\multirow{6}{*}{MNIST} & \LeakyReluFive       & 500                                                                     & 5                                                          & \(\leakyrelu\times5\)                                                                  \\ \cline{2-5}
		                       & \LeakyReluEight      & 800                                                                     & 8                                                          & \(\leakyrelu\times8\)                                                                  \\ \cline{2-5}
		                       & \LeakyReluFourteen   & 392                                                                     & 14                                                         & \(\leakyrelu\times14\)                                                                 \\ \cline{2-5}
		                       & \ReluSignMax         & 511                                                                     & 6                                                          & \(\relu\times4,\sign,\maxxx\)                                                          \\ \cline{2-5}
		                       & \ReluBilinearSoftmax & 586                                                                     & 9                                                          & \(\relu,\bilinear,\softmax\)                                                           \\ \cline{2-5}
		                       & \LeakyReluSigmoid    & 280                                                                     & 10                                                         & \(\begin{matrix}\leakyrelu\times6,\phantom{\quad}\\\sigmoid\times3,\relu\end{matrix}\) \\ \hline
	\end{tabular}
	\label{tab:network-topologies}
\end{table}

\section{Generalized INVPROP}\label{appendix:theorem-two-details}
In this section we present a general formulation of Theorem 2
from~\cite[Appendix C]{SuChZiKrHu23}, which served as a key building
block for the INVPROP algorithm for hyper-plane optimization. Our
theorem supports optimizing hyper-planes based on previously derived
hyper-planes, and it tolerates more general input domains \(\Din\)
and activation functions \(\sig{i}\).

\mysubsection{Theorem.} Given constant vectors
\(\vechc{i}, \vecc{i}\), we seek to optimize the bias \(t\) of the
bounding hyper-plane \(\{\vecxx \,|\, \sum_{i=0}^{L-1} \vechctop{i} \vechx{i} + \sum_{i=1}^{L} \vecctop{i} \vecx{i} \ge t\}\)
by solving the dual of the following LP:
\begin{align*}
	\min_{\vecxx, \vechxx} \quad & \sum_{i=0}^{L-1} \vechctop{i} \vechx{i} + \sum_{i=1}^{L} \vecctop{i} \vecx{i}                 \\
	\text { s.t. } \quad         &
	\vecxx \in \Din ;
	\quad \vechx{0} = \vecxx                                                                                                     \\
	                             & \sum_{i=0}^{L-1} \vechC{i} \vechx{i} + \sum_{i=1}^{L} \vecC{i} \vecx{i} + \vecdd \le \veczero \\
	                             & \vecx{i} = \vecW{i} \vechxx^{(i-1)}+\vecb{i}                                                  \\
	                             & \vecWl{i} \vecx{i} + \vecbl{i} \le \vechx{i} \le \vecWu{i} \vecx{i} + \vecbu{i}
\end{align*}\label{eqn:generalized-invprop-relaxed-lp}

Theorem~\ref{thm:generalized-invprop} lists a lower bound for this
linear program.
\begin{theorem}\label{thm:generalized-invprop}
	For any \(\vecalpha\),
	\(\vecgamma \ge \veczero\), \(g(\vecalpha, \vecgamma)\) is a
	lower bound to the above linear program where \(g\) is defined
	via
	\begin{align*}
		g(\vecalpha, \vecgamma) & = \inf_{\vecxx \in \Din} \left( \vechc{0} - \vecnutop{1} \vecW{1} + \vecgammatop \vechC{0} \right)^{\top} \vecxx                                               \\
		                        & - \sum_{i=1}^{L} \vecnutop{i} \vecb{i} - \sum_{i=1}^{L-1} \left( \mpos{\vechnutop{i}} \vecbu{i} - \mneg{\vechnutop{i}} \vecbl{i} \right) + \vecgammatop \vecdd
	\end{align*}\label{eqn:generalized-invprop-lower-bound}
	where the terms \(\vecnu{i}, \vechnu{i}\) could be computed
	recursively with
	\begin{align*}
		\vecnu{L}    & = - \vecctop{L} - \vecgammatop \vecC{L}                                                                              \\
		\vechnu{i}   & = \vecnutop{i+1} \vecW{i+1} - \vecgammatop \vechC{i} - \vechctop{i}, i \in [L-1]                                     \\
		\vecnutop{i} & = \mpos{\vechnutop{i}} \vecWu{i} - \mneg{\vechnutop{i}} \vecWl{i} - \vecgammatop \vecC{i} - \vecctop{i}, i \in [L-1]
	\end{align*}\label{eqn:generalized-invprop-recursion-formula}
\end{theorem}

\mysubsection{Global Bounds for Common Domains.} Here are multiple
approaches to solving the infimum
\(\inf_{\vecxx \in \Din} \left( \vechc{0} - \vecnutop{1} \vecW{1} + \vecgammatop \vechC{0} \right)^{\top} \vecxx\)
for common input domains \(\Din\).
\begin{itemize}
	\item When \(\Din\) is the hyper-rectangle \(\vecl{0} \le \vecxx \le \vecu{0}\), concretization results in the minimum value \(\bigmpos{\vecc{0} - \vecnutop{1} \vecW{1}}^{\top} \vecu{0} - \bigmneg{\vecc{0} - \vecnutop{1} \vecW{1}}^{\top} \vecl{0}\).
	\item When \(\Din\) is the \(\ell_p\)-ball \(\| \vecxx - \vecxtag \|_{p} \le \varepsilon\) for \(\varepsilon > 0, p \in [1, \infty), \vecxtag \in \R{n_0}\) then, by duality~\cite{ZhWeChHsDa18}, \(-\varepsilon \|\vecc{0} - \vecnutop{1} \vecW{1}\|_{q} + \left(\vecc{0} - \vecnutop{1} \vecW{1}\right)^{\top} \vecxtag\) is no larger than the infimum, where \(\tfrac{1}{p}+\tfrac{1}{q}=1\).
	\item When \(\Din\) is a polyhedron \(\vecA \vecxx + \vecbb \le \veczero\) then the infimum might be calculated by solving the corresponding LP in the input space.
\end{itemize}

\mysubsection{Between Theorem~\ref{thm:generalized-invprop} and INVPROP.}
Our theorem is a general version of Theorem 2
from~\cite[Appendix C]{SuChZiKrHu23}. A ReLU's triangular linear
relaxation~\cite{XuZhWaWaJaLiHs20} is replaced by the relaxations
\(\vecWl{i} \vecx{i} + \vecbl{i} \le \vechx{i} \le \vecWu{i} \vecx{i} + \vecbu{i}\),
the output constraint \(\vecx{L} \le \veczero\) is superseded by the
general bounding polyhedron \(\sum_{i=0}^{L-1}\vecC{i}\vecx{i} + \sum_{i=1}^{L}\vechC{i}\vechx{i} + \vecdd \le \veczero\),
and the first global lower bound is replaced by an infimum
expression. The reader may verify that, by substituting back all
these in Theorem~\ref{thm:generalized-invprop}, then the linear
program, lower bound \(g(\vecalpha, \vecgamma)\) and recursion
formula for \(\vecnuu, \vechnuu\) evaluate to those of Theorem 2
in~\cite[Appendix C]{SuChZiKrHu23}.

\newpage

\section{Proof of Theorem~\ref{thm:generalized-invprop}}\label{appendix:theorem-proof}

Let us establish the correctness of our generalized theorem using a
close argument to\protect{~\cite[Appendix D.]{SuChZiKrHu23}}. We'll
start off by taking the Lagrange of a majority of the LP's
constraints.
\begin{align*}
	\min_{\vecxx, \vechxx} \max_{\vecnuu, \vectauu, \vecpii, \vecgamma, \vecalpha} \quad & \sum_{i=0}^{L-1} \vechctop{i} \vechx{i} + \sum_{i=1}^{L} \vecctop{i} \vecx{i}                                                                       \\
	                                                                                     & + \vecgammatop \left( \sum_{i=0}^{L-1} \vechC{i} \vechx{i} + \sum_{i=1}^{L} \vecC{i} \vecx{i} + \vecdd \right)                                      \\
	                                                                                     & + \sum_{i=1}^{L} \vecnutop{i} \left( \vecx{i} - \vecW{i} \vechx{i-1} - \vecb{i} \right)                                                             \\
	                                                                                     & + \sum_{i=1}^{L-1} \vecpitop{i} \left( \vechx{i} - \vecWu{i} \vecx{i} - \vecbu{i} \right)                                                           \\
	                                                                                     & - \sum_{i=1}^{L-1} \vectautop{i} \left( \vechx{i} - \vecWl{i} \vecx{i} - \vecbl{i} \right)                                                          \\
	\text { s.t. } \quad
	                                                                                     & \vecl{0} \le \vechx{0} \le \vecu{0} ; \quad \vectauu \ge 0 ; \quad \vecpii \ge 0 ; \quad \vecgamma \ge 0
\end{align*}\label{eqn:generalized-invprop-dual}

According to the Strong Duality Theorem, Reversing the optimization
order and re-arranging results in the following equivalent LP:
\begin{align*}
	\max_{\vecnuu, \vectauu, \vecpii, \vecgamma, \vecalpha} \min_{\vecxx, \vechxx} \quad & \left( \vecctop{L} + \vecnutop{L} + \vecgammatop \vecC{L} \right) \vecx{L}                                                                         \\
	                                                                                     & + \left( \vechc{0} - \vecnutop{1} \vecW{1} + \vecgammatop \vechC{0} \right)^{\top} \vechx{0}                                                       \\
	                                                                                     & + \sum_{i=1}^{L-1} \left( \vecctop{i} + \vecnutop{i} - \vecpitop{i} \vecWu{i} + \vectautop{i} \vecWl{i} + \vecgammatop \vecC{i} \right) \vecx{i}   \\
	                                                                                     & + \sum_{i=1}^{L-1} \left( \vechctop{i} - \vecnutop{i+1} \vecW{i+1} + \vecpitop{i} - \vectautop{i} + \vecgammatop \vechC{i} \right) \vechx{i}       \\
	                                                                                     & - \sum_{i=1}^{L} \vecnutop{i} \vecb{i} - \sum_{i=1}^{L-1} \left( \vecpitop{i} \vecbu{i} - \vectautop{i} \vecbl{i} \right) + \vecgammatop \vecdd    \\
	\text { s.t. } \quad
	                                                                                     & \vecl{0} \le \vechx{0} \le \vecu{0}; \quad \vectauu \ge 0 ; \quad \vecpii \ge 0 ; \quad \vecgamma \ge 0
\end{align*}\label{eqn:generalized-invprop-reordered-dual}

To solve the inner optimization, notice the variables
\(\vecx{i}, \vechx{i}\) are unconstrained, meaning that their
coefficients must equal zero else the outer optimization's objective
would be unbounded. Eliminating the inner optimization variables
\(\vecx{i}, \vechx{i}\) results in this LP:
\begin{align*}
	\max_{\vecnuu, \vectauu, \vecpii, \vecgamma, \vecalpha} \quad & \inf_{\vecxx \in \Din} \left( \vechc{0} - \vecnutop{1} \vecW{1} + \vecgammatop \vechC{0} \right)^{\top} \vecxx                                  \\
	                                                              & - \sum_{i=1}^{L} \vecnutop{i} \vecb{i} - \sum_{i=1}^{L-1} \left( \vecpitop{i} \vecbu{i} - \vectautop{i} \vecbl{i} \right) + \vecgammatop \vecdd \\
	\text { s.t. } \quad
	                                                              & \vecnu{L} = - \vecctop{L} - \vecgammatop \vecC{L}                                                                                               \\
	                                                              & \vecnutop{i} = \vecpitop{i} \vecWu{i} - \vectautop{i} \vecWl{i} - \vecctop{i} - \vecgammatop \vecC{i}, i \in [L-1]                              \\
	                                                              & \vecnutop{i+1} \vecW{i+1} - \vecgammatop \vechC{i} - \vechctop{i} = \vecpitop{i} - \vectautop{i}, i \in [L-1]                                   \\
	                                                              & \vectauu \ge 0 ; \quad \vecpii \ge 0 ; \quad \vecgamma \ge 0
\end{align*}\label{eqn:generalized-invprop-semi-solved-dual}

Denote
\(\vechnu{i} = \vecnutop{i+1} \vecW{i+1} - \vecgammatop \vechC{i} - \vechctop{i}\).
The constraints \(\vecpii \ge 0\), \(\vectauu \ge 0\) and
\(\vecnutop{i+1} \vecW{i+1} - \vecgammatop \vechC{i} = \vecpitop{i} - \vectautop{i}\)
imply setting the values of \(\vecpi{i}\) and \(\vectau{i}\) to be
\(\vecpi{i} = \mpos{\vechnu{i}}\) and
\(\vectau{i} = \mneg{\vechnu{i}}\) yields a valid lower bound for the
optimization. Combined with the two other equality constraints of the
linear program, we arrive at the following recursive relation for
\(\vecnu{i}, \vechnu{i}\):
\begin{align*}
	\vecnu{L}    & = - \vecctop{L} - \vecgammatop \vecC{L}                                                                              \\
	\vechnu{i}   & = \vecnutop{i+1} \vecW{i+1} - \vecgammatop \vechC{i} - \vechctop{i}, i \in [L-1]                                     \\
	\vecnutop{i} & = \mpos{\vechnutop{i}} \vecWu{i} - \mneg{\vechnutop{i}} \vecWl{i} - \vecgammatop \vecC{i} - \vecctop{i}, i \in [L-1]
\end{align*}\label{eqn:generalized-invprop-derived-recursion-formula}

Overall, we have that the optimal value for the original linear
program is no smaller than the solution of this LP:
\begin{align*}
	\max_{\vecgamma, \vecalpha} \quad & \inf_{\vecxx \in \Din} \left( \vechc{0} - \vecnutop{1} \vecW{1} + \vecgammatop \vechC{0} \right)^{\top} \vecxx                                                 \\
	                                  & - \sum_{i=1}^{L} \vecnutop{i} \vecb{i} - \sum_{i=1}^{L-1} \left( \mpos{\vechnutop{i}} \vecbu{i} - \mneg{\vechnutop{i}} \vecbl{i} \right) + \vecgammatop \vecdd \\
	\text { s.t. } \quad
	                                  & \vecgamma \ge 0
\end{align*}\label{eqn:generalized-invprop-derived-lower-bound}

Consequently, \(g(\vecalpha, \vecgamma)\) is a lower bound to the
original LP's solution for every choice of \(\vecgamma \ge 0\) and
\(\vecalpha\), where \(g\) is the objective function of the above
optimization problem.

\newpage

\section{Running Example Details}\label{appendix:running-example-details}

In this section we include additional computations regarding the
running example in
Subsection~\ref{instantiating-pmnr:subsec:running-example}.

\begin{figure}[htb]
	\centering
	\begin{tikzpicture}[>=Stealth,
			node distance=10ex,
			text height=1.5ex,text depth=.25ex,
			invis/.style={circle, minimum size=0ex},
			input/.style={circle, minimum size=5ex,
					thick, draw=blue, fill=white},
			hidden/.style={circle, minimum size=5ex,
					thick, draw=black!50!white, fill=white},
			output/.style={circle, minimum size=5ex,
					thick, draw=red, fill=white},
			background/.style={rounded corners,
			           fill=blue!10!white, draw=black}]

		\node (x0_0)   [input]                                     {$\hx{0}{0}$};
		\node (x0_1)   [input, below=15ex of x0_0]                 {$\hx{0}{1}$};
		\node (x1_0)   [hidden, right=15ex of x0_0, yshift=10ex]   {$\hx{1}{0}$};
		\node (x1_1)   [hidden, right=15ex of x0_0, yshift=-10ex]  {$\hx{1}{1}$};
		\node (x1_2)   [hidden, right=15ex of x0_1, yshift=-10ex]  {$\hx{1}{2}$};
		\node (x2_0)   [hidden, right=15ex of x1_0, yshift=-10ex]  {$\hx{2}{0}$};
		\node (x2_1)   [hidden, right=15ex of x1_2, yshift=10ex]   {$\hx{2}{1}$};
		\node (x3_0)   [output, right=15ex of x2_0, yshift=-10ex]  {$\x{3}{0}$};

		\draw [->] (x0_0) -- (x1_0)    node[midway, above]                         {1};
		\draw [->] (x0_0) -- (x1_1)    node[midway, below]                         {2};
		\draw [->] (x0_1) -- (x1_1)    node[midway, above]                         {-3};
		\draw [->] (x0_1) -- (x1_2)    node[midway, below]                         {1};

		\draw [->] (x1_0) -- (x2_0)    node[midway, above]                         {1};
		\draw [->] (x1_0) -- (x2_1)    node[near start, above right, yshift=-1ex]  {-1};
		\draw [->] (x1_1) -- (x2_0)    node[very near start, above]                {1};
		\draw [->] (x1_1) -- (x2_1)    node[very near start, below]                {1};
		\draw [->] (x1_2) -- (x2_0)    node[near start, below right, yshift=1ex]   {-1};
		\draw [->] (x1_2) -- (x2_1)    node[midway, below]                         {-5};

		\draw [->] (x2_0) -- (x3_0)   node[midway, above]                          {-1};
		\draw [->] (x2_1) -- (x3_0)   node[midway, below]                          {-3};

		\node (x1_0_bias)   [align=center, below=0.5ex of x1_0]   {1};
		\node (x1_1_bias)   [align=center, below=0.5ex of x1_1]   {0};
		\node (x1_2_bias)   [align=center, below=0.5ex of x1_2]   {0};

		\node (x2_0_bias)   [align=center, below=0.5ex of x2_0]   {0};
		\node (x2_1_bias)   [align=center, below=0.5ex of x2_1]   {2};

		\node (x3_0_bias)   [align=center, below=0.5ex of x3_0]   {26.1};

		\node (x0_0_range)   [align=center, above=0.5ex of x0_0]  {$[-1, 1]$};
		\node (x0_1_range)   [align=center, above=0.5ex of x0_1]  {$[-1, 1]$};

		\node (x1_0_range)   [align=center, above=0.5ex of x1_0]  {$[0, 2]$};
		\node (x1_1_range)   [align=center, above=0.5ex of x1_1]  {$[0, 5]$};
		\node (x1_2_range)   [align=center, above=0.5ex of x1_2]  {$[0, 1]$};

		\node (x2_0_range)   [align=center, above=0.5ex of x2_0]  {$[0, 4]$};
		\node (x2_1_range)   [align=center, above=0.5ex of x2_1]  {$[0, 6]$};

		\node (x3_0_range)   [align=center, above=0.5ex of x3_0]  {$[12.1, 26.1]$};

		\node(hx0_0_deeppoly) [align=left, below=42.5ex of x0_0]  {$-1 \le \hx{0}{0} \le 1$\\$\hx{0}{0} \in [-1, 1]$};
		\node(hx0_1_deeppoly) [align=left, below=5ex of hx0_0_deeppoly]  {$-1 \le \hx{0}{1} \le 1$\\$\hx{0}{1} \in [-1, 1]$};

		\node(x1_0_deeppoly)  [align=left, below=5ex of hx0_1_deeppoly]  {$\x{1}{0} = \hx{0}{0} + 1$\\$\x{1}{0} \in [0, 2]$};
		\node(hx1_0_deeppoly) [align=left, below=5ex of x1_0_deeppoly]  {$\x{1}{0} \le \hx{1}{0} \le \x{1}{0}$\\$\hx{1}{0} \in [0, 2]$};

		\node(x1_1_deeppoly)  [align=left, right=0ex of hx0_0_deeppoly, xshift=2.25ex]  {$\x{1}{1} = 2\hx{0}{0} - 3\hx{0}{1}$\\$\x{1}{1} \in [-5, 5]$};
		\node(hx1_1_deeppoly) [align=left, below=5ex of x1_1_deeppoly, xshift=-3.75ex]  {$0 \le \hx{1}{1} \le 5$\\$\hx{1}{1} \in [0, 5]$};

		\node(x1_2_deeppoly)  [align=left, below=5ex of hx1_1_deeppoly, xshift=0.5ex]  {$\x{1}{2} = \hx{0}{1}$\\$\x{1}{2} \in [-1, 1]$};
		\node(hx1_2_deeppoly) [align=left, below=5ex of x1_2_deeppoly, xshift=-0.5ex]  {$0 \le \hx{1}{2} \le 1$\\$\hx{1}{2} \in [0, 1]$};

		\node(x2_0_deeppoly)  [align=left, right=0ex of x1_1_deeppoly]  {$\x{2}{0} = \hx{1}{0} + \hx{1}{1} - \hx{1}{2}$\\$\x{2}{0} \in [-1, 7]$};
		\node(hx2_0_deeppoly) [align=left, below=5ex of x2_0_deeppoly, xshift=-0.25ex]  {$\x{2}{0} \le \hx{2}{0} \le \tfrac{7}{8}\x{2}{0}+\tfrac{7}{8}$\\$\hx{2}{0} \in [-1, 7]$};

		\node(x2_1_deeppoly)  [align=left, below=5ex of hx2_0_deeppoly, xshift=2.25ex]  {$\x{2}{1} = 2-\hx{1}{0} + \hx{1}{1}- 5\hx{1}{2}$\\$\x{2}{1} \in [-5, 7]$};
		\node(hx2_1_deeppoly) [align=left, below=5ex of x2_1_deeppoly, xshift=-1.25ex]  {$\x{2}{1} \le \hx{2}{1} \le \tfrac{7}{12}\x{2}{1}+\tfrac{35}{12}$\\$\hx{2}{1} \in [-5, 7]$};

		\node(x3_0_deeppoly)  [align=left, right=2ex of x2_0_deeppoly, xshift=-1ex, yshift=-4.5ex]  {$\begin{matrix}\x{3}{0} = -\hx{2}{0}\\ -3\hx{2}{1} + 26.1\end{matrix}$\\$\begin{matrix}\x{3}{0} \in\phantom{\qquad\quad}\\\mathbf{[-0.15, 40.1]}\end{matrix}$};

		\node(alg_name_deeppoly)  [invis, below=12ex of x3_0_deeppoly] {};

		\node (background_adjustment_deeppoly) [invis, above=2ex of hx0_0_deeppoly] {};
		\begin{scope}[on background layer]
			\node (background_deeppoly) [background,fit=(alg_name_deeppoly)(background_adjustment_deeppoly)(hx0_0_deeppoly)(hx0_1_deeppoly)(x1_0_deeppoly)(x1_1_deeppoly)(x1_2_deeppoly)(hx1_0_deeppoly)(hx1_1_deeppoly)(hx1_2_deeppoly)(x2_0_deeppoly)(x2_1_deeppoly)(hx2_0_deeppoly)(hx2_1_deeppoly)(x3_0_deeppoly)] {};
		\end{scope}

	\end{tikzpicture}
	\caption{The DNN shown in Fig.~\ref{fig:example-network} and
	         DeepPoly bounds for it given \(\Din=[-1, 1]^2\).}
	\label{fig:example-network-deep-poly}
\end{figure}
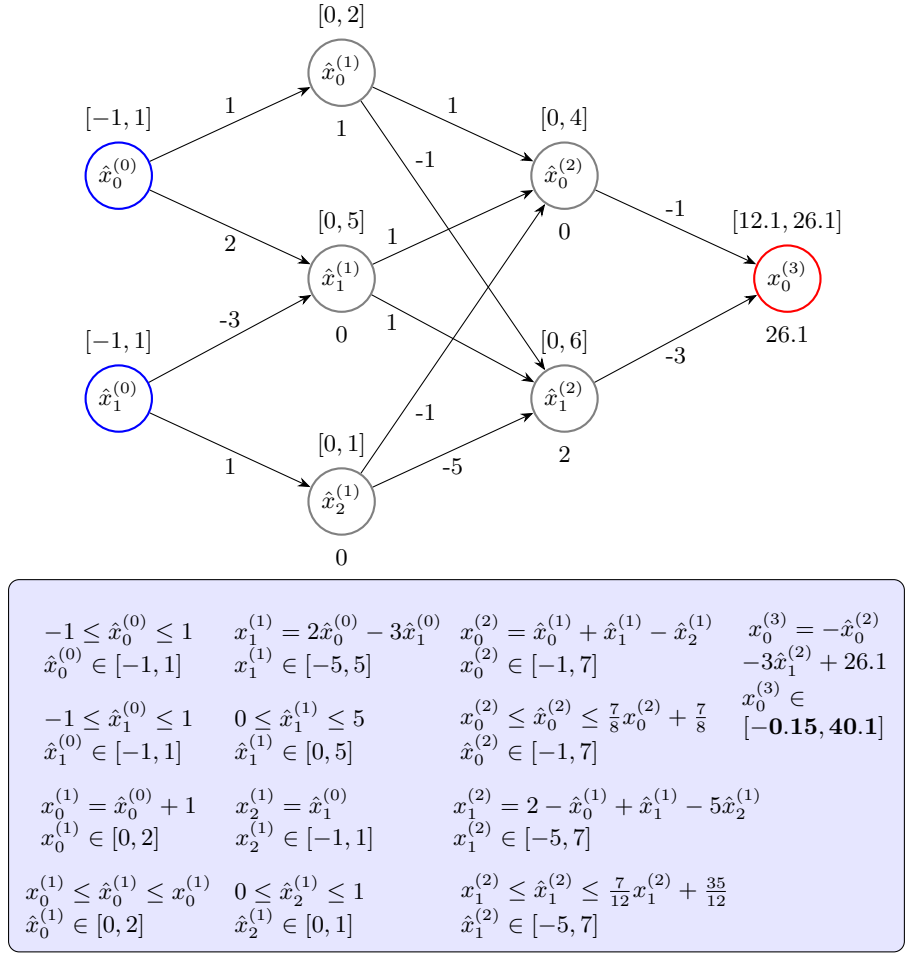

\mysubsection{DeepPoly Outputs.}
Fig.~\ref{fig:example-network-deep-poly} shows the linear
over-approximations and concrete bounds produced by the Single-Neuron
Relaxation-based DeepPoly.

\mysubsection{Hyper-planes learned by PMNR-ALL.} Here are the \(12\)
hyper-planes generated by \(\pmnrall\) (in an arbitrary order) by
invoking BHSO and PGD for optimization of the lower bound listed in
Theorem~\ref{thm:generalized-invprop}:
{\allowdisplaybreaks
\begin{align*}
	 \hx{1}{1} + \hx{1}{2}                                                      &\le 6,    \\
	 - \hx{1}{1} - \hx{1}{2}                                                    &\le 0,    \\
	 - \hx{1}{1} + \hx{1}{2}                                                    &\le 1,    \\
	 \hx{1}{1} - \hx{1}{2}                                                      &\le 5,    \\
	 (\hx{2}{0} - \x{2}{0}) + (\hx{2}{1} - \x{2}{1})                            &\le 5,    \\
	 - (\hx{2}{0} - \tfrac{7}{8}\x{2}{0}) - (\hx{2}{1} - \tfrac{7}{12}\x{2}{1}) &\le 1.96, \\
	 - (\hx{2}{0} - \x{2}{0}) + (\hx{2}{1} - \x{2}{1})                          &\le 5,    \\
	 (\hx{2}{0} - \tfrac{7}{8}\x{2}{0}) - (\hx{2}{1} - \tfrac{7}{12}\x{2}{1})   &\le 2.96, \\
	 (\hx{2}{0} - \x{2}{0}) - (\hx{2}{1} - \x{2}{1})                            &\le 1,    \\
	 (\hx{2}{0} - \tfrac{7}{8}\x{2}{0}) + (\hx{2}{1} - \tfrac{7}{12}\x{2}{1})   &\le 3.04, \\
	 - (\hx{2}{0} - \x{2}{0}) - (\hx{2}{1} - \x{2}{1})                          &\le 0,    \\
	 (\hx{2}{0} - \tfrac{7}{8}\x{2}{0}) + (\hx{2}{1} - \tfrac{7}{12}\x{2}{1})   &\le 3.54
\end{align*}\label{eqn:example-network-pmnr-all}
}

\mysubsection{Hyper-planes learned by PMNR.} Below are the \(8\)
multi-neuron bounds \(\pmnr\) inferred in an identical manner to
PMNR-ALL. They are slightly looser than the relaxations produced by
PMNR-ALL.
{\allowdisplaybreaks
\begin{align*}
	 (\hx{2}{0} - \x{2}{0}) + (\hx{2}{1} - \x{2}{1})                            &\le 20,   \\
	 - (\hx{2}{0} - \tfrac{7}{8}\x{2}{0}) - (\hx{2}{1} - \tfrac{7}{12}\x{2}{1}) &\le 2.46, \\
	 - (\hx{2}{0} - \x{2}{0}) + (\hx{2}{1} - \x{2}{1})                          &\le 5.14, \\
	 (\hx{2}{0} - \tfrac{7}{8}\x{2}{0}) - (\hx{2}{1} - \tfrac{7}{12}\x{2}{1})   &\le 2.99, \\
	 (\hx{2}{0} - \x{2}{0}) - (\hx{2}{1} - \x{2}{1})                            &\le 1.02, \\
	 (\hx{2}{0} - \tfrac{7}{8}\x{2}{0}) + (\hx{2}{1} - \tfrac{7}{12}\x{2}{1})   &\le 3.07, \\
	 - (\hx{2}{0} - \x{2}{0}) - (\hx{2}{1} - \x{2}{1})                          &\le 0,    \\
	 (\hx{2}{0} - \tfrac{7}{8}\x{2}{0}) + (\hx{2}{1} - \tfrac{7}{12}\x{2}{1})   &\le 3.54
\end{align*}\label{eqn:example-network-pmnr}
}

\newpage

\section{Evaluation Within Marabou}\label{appendix:evaluation-marabou}

\subsection{Neuron Selection Heuristic}\label{evaluation-marabou:subsec:neuron-selection-heuristic}
To quantify the impact of our proposed NSSE heuristic, we implemented
another instantiation of the PMNR, named PMNR-Random, which only
differs from the main instantiation \(\pmnr\) in terms of its neuron
selection heuristic. As opposed to \(\pmnr\) which chooses neurons
with maximal NSSE score, PMNR-Random picks a layer \(\ell\) uniformly
at random, then proceeds to selects \(d\) unfixed-phase neurons from
it. In an identical fashion to our experiments in
Section~\ref{sec:experiments-and-evaluation}, we replaced the initial
bound tightening algorithm of Marabou with PMNR-Random and compared
it to PMNR-enhanced Marabou on \(344\) local robustness queries.

\begin{figure}[htb]
	\includegraphics[width=\linewidth,trim=0ex 0ex 0ex 0.5ex,clip]{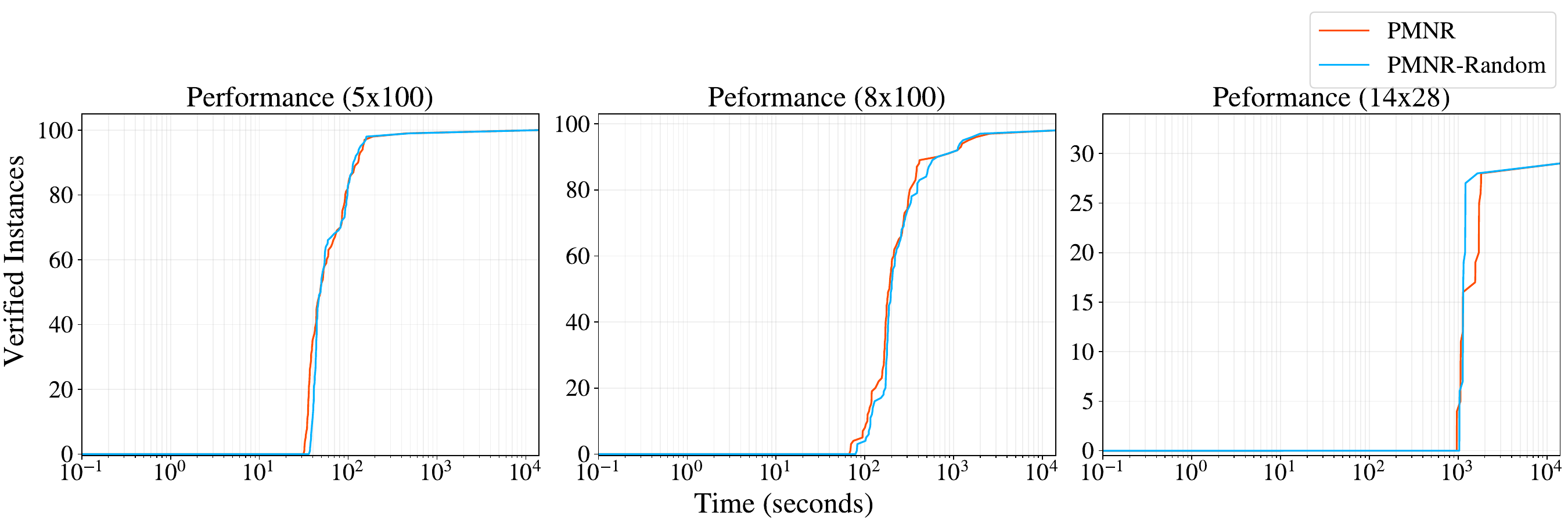}
	\includegraphics[width=\linewidth,trim=0ex 0ex 0ex 0.5ex,clip]{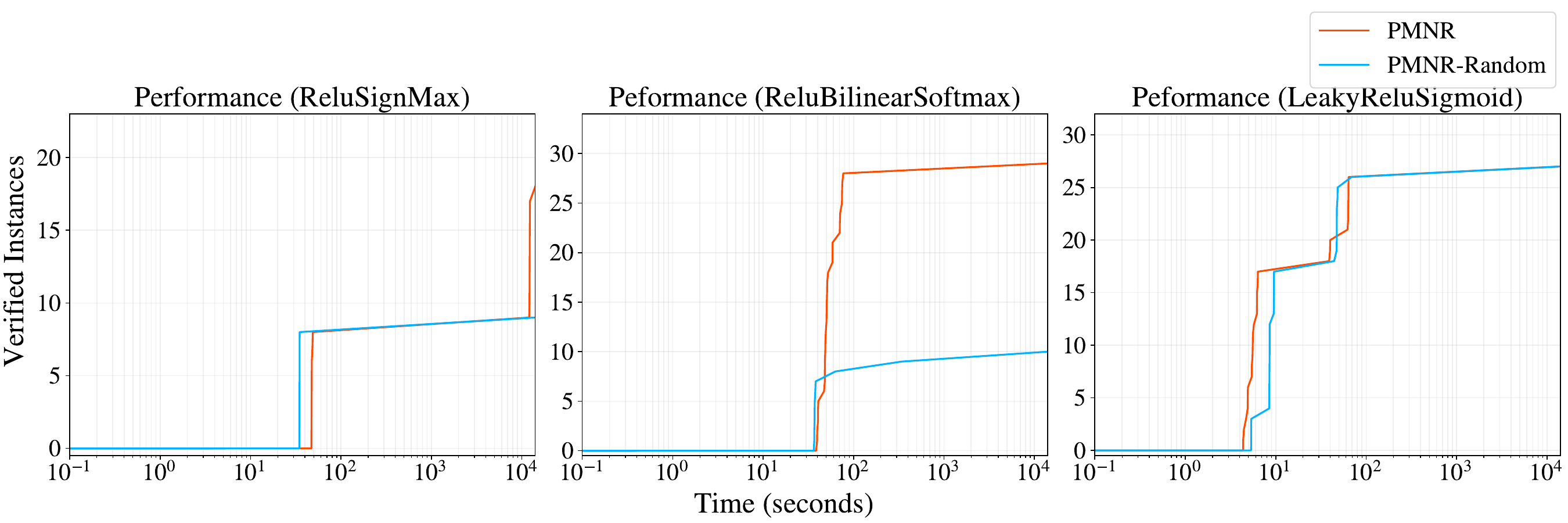}
	\caption{Cumulative instances verified with PMNR-enhanced Marabou
	and PMNR-Random-enhanced Marabou, versus time (seconds, log scale)
	required for verification, for every benchmark class.}\label{fig:cactus-plot-neuron-selection}
\end{figure}

Aggregate results are shown at
Table~\ref{tab:results-neuron-selection} and Marabou's cumulative
runtime per model is depicted in
Fig.~\ref{fig:cactus-plot-neuron-selection}. Overall, leveraging the
NSSE heuristic within PMNR-enhanced Marabou resulted in \(301\)
queries being verified out of \(344\), an \(10\%\) improvement over
random neuron selection (\(273\) verified). The largest gains were
noted for the piecewise-linear \(\ReluSignMax\) and the
non-piecewise-linear \(\ReluBilinearSoftmax\) DNNs, for which
NSSE-based neuron selection has led to a \(\times2\)-\(2.9\) more
verified queries (\(27\) and \(29\) solved queries with NSSE, in
contrast to \(9\) and \(10\) solved queries without NSSE,
respectively).

\begin{table*}[htb]
    \centering
	\caption{Comparing NSSE-based to random neuron selection in \(\pmnr\).}
	\begin{tabular}{|c|c|cc|cc|}
		\hline
		\multirow{2}{*}{Model}                   & \multirow{2}{*}{Queries}       &
		\multicolumn{2}{c|}{\(\pmnr\) (NSSE)} & \multicolumn{2}{c|}{\(\pmnr\) (Random)}                                                                                                                             \\ \cline{3-6}
		                                         &                                & \multicolumn{1}{c|}{Solved \(\uparrow\)} & Time \(\downarrow\) & \multicolumn{1}{c|}{Solved \(\uparrow\)} & Time \(\downarrow\) \\ \hline
		\LeakyReluFive                           & 100                            & \multicolumn{1}{c|}{\textbf{100}}        & \textbf{67}         & \multicolumn{1}{c|}{100}                 & 68                  \\ \hline
		\LeakyReluEight                          & 100                            & \multicolumn{1}{c|}{\textbf{98}}         & \textbf{294}        & \multicolumn{1}{c|}{98}                  & 305                 \\ \hline
		\LeakyReluFourteen                       & 36                             & \multicolumn{1}{c|}{\textbf{29}}         & 1321                & \multicolumn{1}{c|}{\textbf{29}}         & \textbf{1147}       \\ \hline
		\ReluSignMax                             & 36                             & \multicolumn{1}{c|}{\textbf{18}}         & 6138                & \multicolumn{1}{c|}{9}                   & \textbf{34}         \\ \hline
		\ReluBilinearSoftmax                     & 36                             & \multicolumn{1}{c|}{\textbf{29}}         & \textbf{54}         & \multicolumn{1}{c|}{10}                  & 69                  \\ \hline
		\LeakyReluSigmoid                        & 36                             & \multicolumn{1}{c|}{\textbf{27}}         & 22                  & \multicolumn{1}{c|}{27}                  & \textbf{21}         \\ \hline
		Total                                    & 344                            & \multicolumn{1}{c|}{\textbf{301}}        & 619                 & \multicolumn{1}{c|}{273}                 & \textbf{261}        \\ \hline
	\end{tabular}
	\label{tab:results-neuron-selection}
\end{table*}

\subsection{Stop Condition}\label{evaluation-marabou:subsec:stop-condition}
Here we assess the costs and benefits of different \(n\) values for
the stop condition in
Subsection~\ref{instantiating-pmnr:subsec:other-heuristics}. Deciding
on a value for \(n\) presents a dilemma between tightness and
accuracy, since the more iterations the main loop of
Algorithm~\ref{alg:pmnr} completes, the bounds returned from it are
tighter, at the cost of a requiring greater computational resources
to derive. We evaluated the performance of PMNR-enhanced Marabou with
\(n=1\) and compared it to the \(n=10\), the latter being the setting
of choice for all our other experiments in this paper.

\begin{figure}[htb]
	\includegraphics[width=\linewidth,trim=0ex 0ex 0ex 0.5ex,clip]{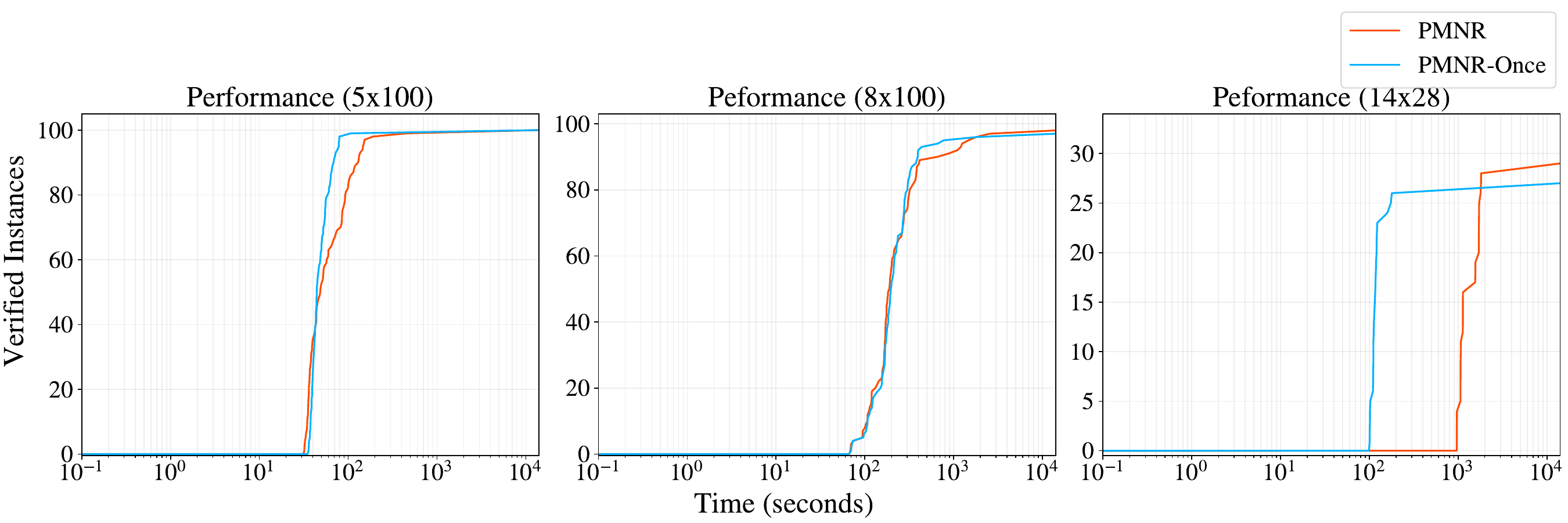}
	\includegraphics[width=\linewidth,trim=0ex 0ex 0ex 0.5ex,clip]{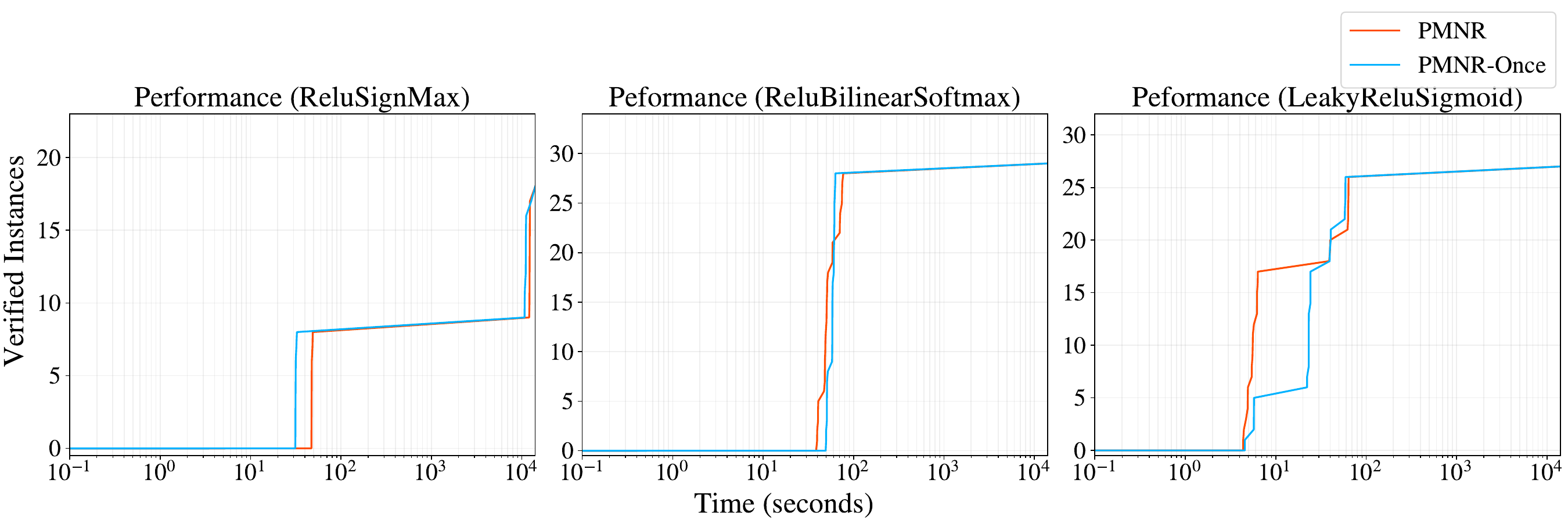}
	\caption{Cumulative instances solved by PMNR-enhanced Marabou (for \(n\) values of \(1\) and \(10\)) versus
	time requirements (seconds, log scale).}\label{fig:cactus-plot-once}
\end{figure}

Aggregate results are shown at Table~\ref{tab:results-once} and
Marabou's cumulative runtime per model is depicted in
Fig.~\ref{fig:cactus-plot-once}. Overall, our default choice of
\(n=10\) has led to three additional queries being verified (\(301\)
versus \(298\), a \(1\%\) gain), albeit with a runtime overhead of
37\% (619 versus 451 seconds). The stronger hyper-planes inferred by
PMNR using \(n=10\) has only impacted the final number of solved
queries for the two piecewise-linear benchmarks \LeakyReluEight,
\LeakyReluFourteen, and the remaining four models have seen no change
in the number of verified queries.

\begin{table*}[htb]
    \centering
	\caption{Comparing maximum main loop iteration values of \(n=1\) and \(n=10\) for \(\pmnr\).}
	\begin{tabular}{|c|c|cc|cc|}
		\hline
		\multirow{2}{*}{Model}                   & \multirow{2}{*}{Queries}       &
		\multicolumn{2}{c|}{\(\pmnr\) (n=10)} & \multicolumn{2}{c|}{\(\pmnr\) (n=1)}                                                                                                                                \\ \cline{3-6}
		                                         &                                & \multicolumn{1}{c|}{Solved \(\uparrow\)} & Time \(\downarrow\) & \multicolumn{1}{c|}{Solved \(\uparrow\)} & Time \(\downarrow\) \\ \hline
		\LeakyReluFive                           & 100                            & \multicolumn{1}{c|}{\textbf{100}}        & 67                  & \multicolumn{1}{c|}{\textbf{100}}        & \textbf{49}         \\ \hline
		\LeakyReluEight                          & 100                            & \multicolumn{1}{c|}{\textbf{98}}         & 294                 & \multicolumn{1}{c|}{97}                  & \textbf{234}        \\ \hline
		\LeakyReluFourteen                       & 36                             & \multicolumn{1}{c|}{\textbf{29}}         & 1321                & \multicolumn{1}{c|}{27}                  & \textbf{137}        \\ \hline
		\ReluSignMax                             & 36                             & \multicolumn{1}{c|}{\textbf{18}}         & 6138                & \multicolumn{1}{c|}{\textbf{18}}         & \textbf{5607}       \\ \hline
		\ReluBilinearSoftmax                     & 36                             & \multicolumn{1}{c|}{\textbf{29}}         & \textbf{54}         & \multicolumn{1}{c|}{\textbf{29}}         & 57                  \\ \hline
		\LeakyReluSigmoid                        & 36                             & \multicolumn{1}{c|}{\textbf{27}}         & \textbf{22}         & \multicolumn{1}{c|}{\textbf{27}}         & 28                  \\ \hline
		Total                                    & 344                            & \multicolumn{1}{c|}{\textbf{301}}        & 619                 & \multicolumn{1}{c|}{298}                 & \textbf{451}        \\ \hline
	\end{tabular}
	\label{tab:results-once}
\end{table*}

\subsection{Comparing PMNR to Forward-Backward Abstract Interpretation}\label{evaluation-marabou:subsec:forward-backward}
In order to further highlight the advantage of the PMNR paradigm over
Single-Neuron Relaxation, we have elected to implement the \FBC
configuration of Forward-Backward Abstract
Interpretation~\cite{WuBaShNaSi22}, using the stop condition laid out
in Subsection~\ref{instantiating-pmnr:subsec:other-heuristics}. We
replaced PMNR with \FBC as Marabou's initial bound tightening method
and analyzed the results.

\begin{figure}[htb]
	\includegraphics[width=\linewidth,trim=0ex 0ex 0ex 0.5ex,clip]{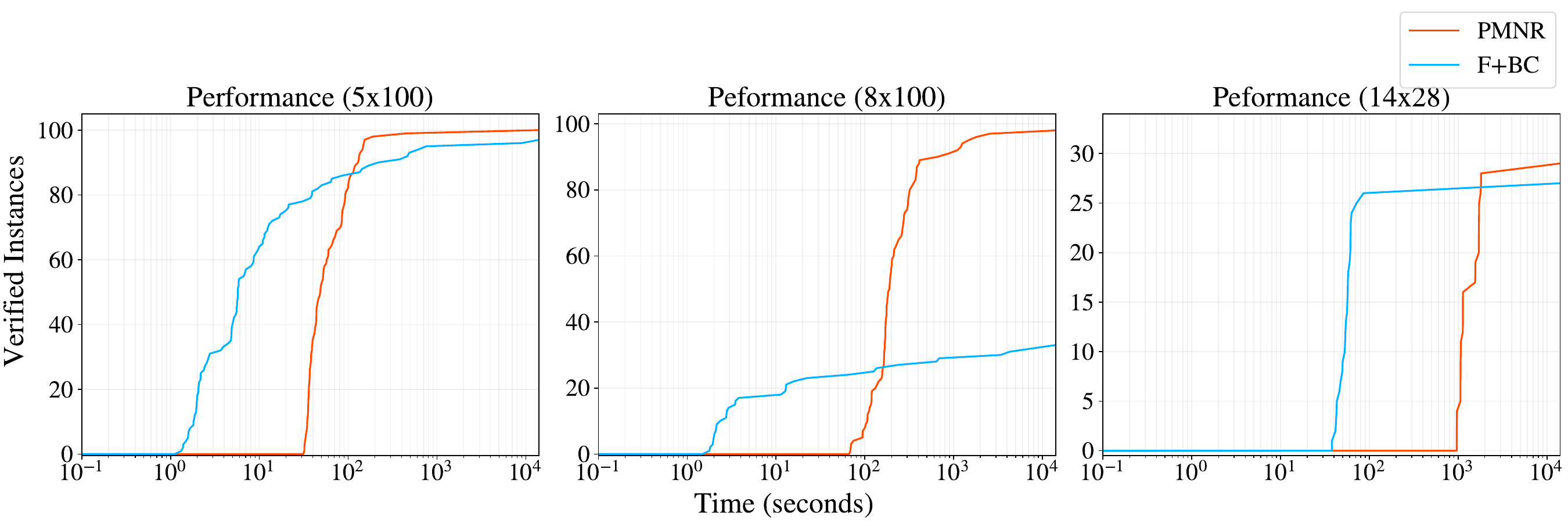}
	\includegraphics[width=\linewidth,trim=0ex 0ex 0ex 0.5ex,clip]{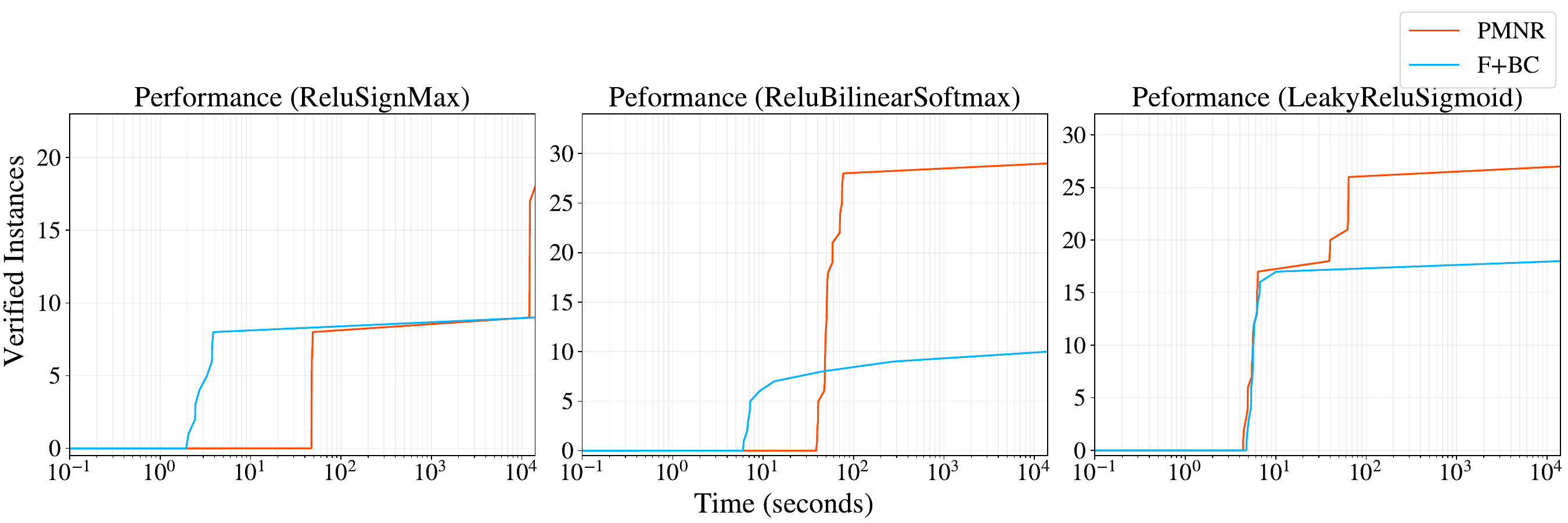}
	\caption{Cumulative instances solved by PMNR-enhanced and
	         Forward-Backward-Analysis-enhanced Marabou versus time
			 costs (seconds, log scale).}\label{fig:cactus-plot-forward-backward}
\end{figure}

Summary statistics are specified at
Table~\ref{tab:results-forward-backward} and Marabou's cumulative
runtime is displayed in Fig.~\ref{fig:cactus-plot-forward-backward}.
These results demonstrate the benefits provided by the tighter bound
inferred by PMNR, as using \(\pmnr\) over \FBC for initial bound
derivation caused a 55\% (\(\tfrac{301}{194}\)) rise in the number of
solved instances, at the cost of a \(\times 3.21\) slower runtime
(619 versus 193 seconds).

Identically to the comparison between DeepPoly and PMNR in
Section~\ref{sec:experiments-and-evaluation}, for all models except
\(\LeakyReluSigmoid\) it holds that the Single-Neuron
Relaxation-based \FBC successfully solves a portion of all queries
faster than PMNR does (due to the additional computing power needed
for multi-neuron bounds), yet PMNR solves the remaining instances
more rapidly compared to \FBC thanks to the tighter relaxations
derived. This is illustrated clearly in
Fig.~\ref{fig:cactus-plot-forward-backward}, where PMNR's cumulative
runtime graph eventually surpasses and towers over \texttt{F+BC}'s
graph.

\begin{table*}[htb]
    \centering
	\caption{Comparing \(\pmnr\) to the \(\FBC\) configuration of Forward-Backward Abstract Interpretation.}
	\begin{tabular}{|c|c|cc|cc|}
		\hline
		\multirow{2}{*}{Model}                   & \multirow{2}{*}{Queries}       &
		\multicolumn{2}{c|}{\(\pmnr\)} & \multicolumn{2}{c|}{\texttt{F+BC}}                                                                                                                                         \\ \cline{3-6}
		                                         &                                & \multicolumn{1}{c|}{Solved \(\uparrow\)} & Time \(\downarrow\) & \multicolumn{1}{c|}{Solved \(\uparrow\)} & Time \(\downarrow\) \\ \hline
		\LeakyReluFive                           & 100                            & \multicolumn{1}{c|}{\textbf{100}}        & \textbf{67}         & \multicolumn{1}{c|}{97}                  & 137                 \\ \hline
		\LeakyReluEight                          & 100                            & \multicolumn{1}{c|}{\textbf{98}}         & \textbf{294}        & \multicolumn{1}{c|}{33}                  & 530                 \\ \hline
		\LeakyReluFourteen                       & 36                             & \multicolumn{1}{c|}{\textbf{29}}         & 1321                & \multicolumn{1}{c|}{27}                  & \textbf{54}         \\ \hline
		\ReluSignMax                             & 36                             & \multicolumn{1}{c|}{\textbf{18}}         & 6138                & \multicolumn{1}{c|}{9}                   & \textbf{2}          \\ \hline
		\ReluBilinearSoftmax                     & 36                             & \multicolumn{1}{c|}{\textbf{29}}         & 54                  & \multicolumn{1}{c|}{10}                  & \textbf{38}         \\ \hline
		\LeakyReluSigmoid                        & 36                             & \multicolumn{1}{c|}{\textbf{27}}         & 22                  & \multicolumn{1}{c|}{18}                  & \textbf{5}          \\ \hline
		Total                                    & 344                            & \multicolumn{1}{c|}{\textbf{301}}        & 619                 & \multicolumn{1}{c|}{194}                 & \textbf{168}        \\ \hline
	\end{tabular}
	\label{tab:results-forward-backward}
\end{table*}

\end{document}